# Exploring Quantum Materials & Applications: A Review

Rajat Kumar Goyal, [a, b]

[a] Interdisciplinary Research Division, Indian Institute of Technology, Karwar, Jheepasani, 342030, Jodhpur, Rajasthan, India

[b] Department of Physics and Astronomy, National Institute of Technology Rourkela 769008, Odisha, India

**Abstract**

Current condensed matter research is centered on advanced materials and their distinctive features. The interest in Quantum materials (QMs) continues to increase without any decrease due to their novel phenomenon and potential as platforms for revolutionary new technologies in modern science and technology. This article emphasizes the exploration of diverse devices and applications facilitated by the unique properties of QMs. Encompassing fields like quantum computing, metrology, sensing, energy, and communication, the review highlights their transformative potential. In QMs, the emerging phenomena are governed by quantum confinement, strong electronic correlations, topology, and symmetry, which makes these materials apart, making them exceptional in their own regard. This paper emphasizes their unique properties, different types of QMs, various interdisciplinary applications, and integration with existing technologies. This study provides a concise overview of diverse discoveries and advancements, presenting a prospective outlook on QMs in multiple domains.

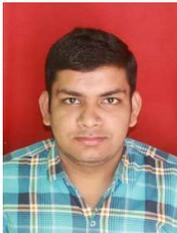

**Rajat Kumar Goyal** received his BSc in Physical sciences from the Ramjas College, University of Delhi, India, and MSc. in Physics from the Department of Physics and Astronomy, National Institute of Technology, Rourkela, India, and he is pursuing an MTech. Degree in Quantum Technology focused on Quantum materials, devices, and sensors from an interdisciplinary research platform (IDRP), Indian Institute of Technology Jodhpur, India. During His Master's, He qualified for national level exams CSIR JRF + NET and GATE. His research interests include an experimental physics perspective on the latest technology and applications, with various materials synthesis and characterizations, quantum materials, devices and sensors, and semiconductor devices. His work is in a multidisciplinary domain, and he has a background in science and technology.

# 1. Introduction

The 20th century witnessed remarkable progress in Physics, significantly enhancing our understanding of sub-atomic and sub-nuclear phenomena. Concurrently, advancements in technology played a pivotal role in driving global development. Most of our quality-of-life improvements over the past half-century can be attributed to semiconductor-based electronics. These technologies have continually evolved, resulting in faster, smaller, energy-efficient, and more affordable products. Today, the size of the smallest transistor measures just a few nanometres, ushering us into an era where quantum principles will increasingly influence device behaviour. This anticipated and inevitable shift opens unprecedented possibilities once deemed impossible. Many technologies crucial to modern life, including communication, computing, sensing, and measurement, have been redesigned to align with the rules of quantum mechanics. In the last two decades, academia and industry have made substantial investments in researching these technologies, with some already progressing beyond the prototype stage. For instance, photonic, superconducting, and atom-based quantum technologies have demonstrated superiority over their classical counterparts.

Recent advances in our understanding of quantum physics in materials have led to a profound shift in perspective. While scientists and engineers have long utilized quantum effects in electronic devices, such as optoelectronics and hard-disk drives, the past decade has revealed how subtle quantum phenomena govern the macroscopic behavior of various materials.[1,2] Quantum Materials (QMs) may be broadly defined as materials that exhibit exotic properties that manifest the quantum nature of their constituent electrons, called QMs. QMs are particularly crucial in the field of quantum computation. Numerous scientists believe that QMs hold the potential to bring about a revolution in various domains, including highly efficient and low-data communication, energy harvesting, semiconductors, and artificial intelligence (AI).[3,4] In the future, AI will play a vital role in performing and analyzing numerous calculations like human brain networks and topology.[5–7] To train an artificial brain network, artificial synapses should match the biological data processing characteristics which show dynamic behaviour toward signal response. QMs attracted the research community due to their highly controllable electronic structure and non-linear performances. Hence, QMs are a developing area for intelligent society.[8,9]

First, going ahead, we will discuss QMs. Suppose we discuss the two recently remarkable aspects of quantum mechanics. The first is the topological nature of quantum wavefunctions, and the second is nonlocal entanglement. QMs can be defined as those with novel entanglement or topological properties, that is, materials with entanglement beyond the requirement of Fermi statistics and with topological responses.[10] Phenomena like quantized vortices in superconductors evidence the topological nature of quantum wavefunctions.[11] These vortices arise due to the necessity of a well-defined phase in the superconducting condensate, with the phase's coupling to magnetic flux governed by gauge invariance. Topological invariants, such as the integer winding number dictating the phase's winding around vortices, remain fixed under smooth system changes, impacting various materials beyond superconductors and facilitating phenomena like dissipation less transport and unique quasiparticle excitations. Secondly, the non-local entanglement inherent in specific quantum states has been emphasized, notably in experiments displaying teleportation with widely separated photons. Non-local entanglement, which highlights the inter-connectedness of quantum states, even within large systems of electrons, offers us a new frontier for technological advancements and fundamental physics.[12,13] Entanglement happens in even simple materials, like metals, where the wave functions of many electrons become very closely linked because electrons are fermionic. Materials exhibiting novel entanglement or topological properties beyond Fermi statistics, such as described vortex formation, are categorized as QMs, displaying entanglement and topological responses beyond conventional frameworks. Understanding these quantum phenomena is a scientific pursuit and a pathway to practical applications and innovation.

We cannot imagine any technology without materials. Therefore, development in the field of materials is always trending. In the last few years, Quantum technology has sparked a revolution in scientific and technological advancement; a few significant applications of Quantum technology are represented in Figure 1. With breakthroughs in quantum key distribution, secure communication has reached unprecedented levels, while quantum networks promise to revolutionize global data transmission. Quantum simulators enable us to delve into complex quantum phenomena, accelerating discoveries in various fields. Meanwhile, QMs exhibit astonishing properties, propelling present material science expertise to uncharted territories[14]. Post-quantum cryptography ensures the robustness of digital security in the face of quantum threats. Quantum sensors push the boundaries of precise measurements, while quantum cloud computing holds the potential to unleash unparalleled computational power. Enhanced by quantum memories, this technological potential becomes even more evident, shaping the future of data storage.[15,16] Behind these innovations lies quantum software, a driving force that maximizes the potential of quantum computing, ushering in an era of computation previously thought impossible.[17]

A material whose properties cannot be fully described by the classical behavior of materials and whose properties originate from novel quantum effects are described as QMs.[18,19] Few novel quantum effects are discussed in this paragraph. In classical materials, we can explain the physical performance of the material by particle interaction. Additionally, a low-level quantum treatment towards electrons may be necessary to describe the behavior in detail. For example, ion motion can be classically explained by electrostatic interactions.[20] Still, a quantum mechanics approach can only provide a complete explanation. Classical theory cannot explain many effects that arise in QMs (like quantum fluctuation and spin entanglements).[21] In QMs, complicated electronic states originate from the interactions of the degree of freedom (lattice, charge, orbital, and spin).[22] So, we observed many strange phenomena like superconductivity[23], topological insulator (TIs)[24], spin liquid[25], and quantum hall effect.[26,27] The 1986 discovery of high-temperature superconductivity in copper oxides initiated a significant expansion in QMs research, revealing that macroscopic quantum phenomena could exist beyond extreme conditions.[28] This breakthrough prompted extensive exploration into the influence of Coulomb interactions on conduction electrons, leading to the investigation of diverse materials. Today's QMs include examples such as hydrogen sulfide, transitioning from a foul-smelling gas to a superconductor with record-setting transition temperatures (exceeding 200 K) under high pressure [29] and diamonds, where the entanglement of electronic and nuclear spins at defect centers enables long-lasting quantum coherence even at room temperature, offering promising prospects for quantum technologies.[30,31]

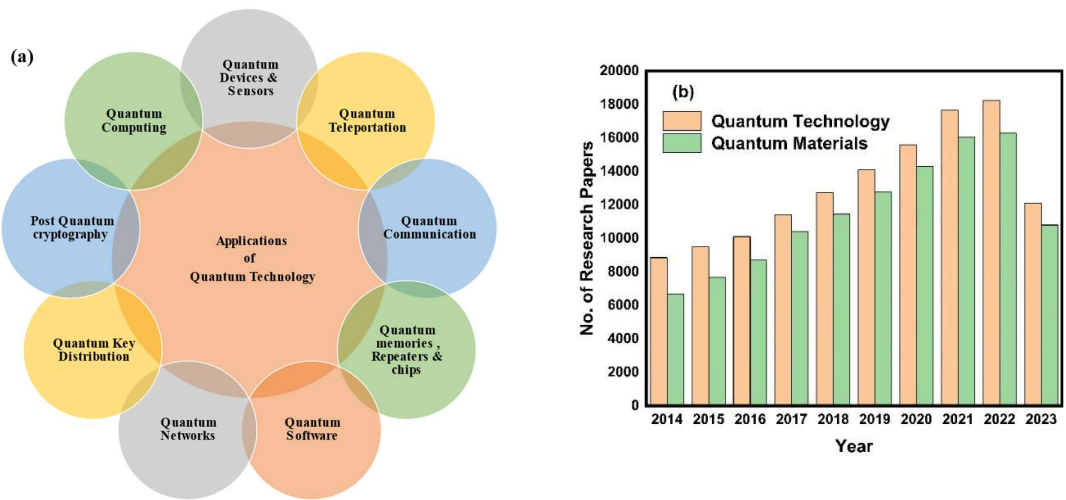

Figure 1: (a) Broad application of Quantum Technology. (b) Number of research articles from 2014 to 2023 in quantum technology and QMs. (Source is web of science).

The 2022 Nobel Prize in Physics was jointly awarded to Alain Aspect, John F. Clauser, and Anton Zeilinger for their exceptional work in quantum physics.[32] They built upon the impressive research of John Stewart Bell,[33] who sought to address questions raised in a 1935 paper by Albert Einstein, Boris Podolsky, and Nathan Rosen. Independently, these three laureates conducted experiments with entangled photons, ultimately proving the violation of Bell inequalities and shedding light on the quantum property of entanglement. This property holds great significance for the development of quantum computers, which have the potential to perform tasks far more efficiently than traditional computers. While quantum mechanics has been a part of physics for over a century and has already led to practical applications like transistors and lasers, its possibilities are virtually boundless. The Nobel Prize committee emphasized that the laureates' innovative experimental tools have paved the way for a new era in quantum technology. The ability to manipulate and control quantum states and their multifaceted properties opens exciting possibilities with far-reaching potential in sensing and communication technologies, making quantum physics a field of ongoing and profound impact on our understanding of the world and developing novel technologies.

The Nobel Prize in Chemistry for 2023 recognizes the innovative discovery and advancement of Quantum Dots (QDs).[34] These QDs, the tiniest components of nanotechnology, have found applications in various fields, including lighting technology, medical surgery, and biological mapping. Usually, an element's properties are dictated by its number of electrons. Still, when the matter is reduced to nanoscale dimensions, quantum phenomena come into play, governed by size rather than electron count. Alexey Ekimov discovered the semiconductor nanocrystals known as QDs in 1981 and explored size-dependent quantum effects in nanoparticles using copper chloride nanoparticles.[35] Louis Brus further advanced the field by proving size-dependent quantum effects in freely suspended particles a few years later. In 1993, Moungi Bawendi's contributions produced almost perfect QDs, a crucial step for their practical applications.[36] Today, QDs illuminate screens in QLED technology, enhance the lighting of LED lamps, and aid biochemists and doctors in tissue mapping. These tiny particles hold immense promise, potentially impacting flexible electronics, miniature sensors, thinner solar cells, and encrypted quantum communication, signifying just the beginning of their transformative potential.

In Figure 1(b), a comprehensive investigation was conducted by carefully analyzing research papers systematically sourced from the Web of Science database, employing the strategic utilization of keywords 'Quantum Technology' and 'QMs.' This analytical endeavor, spanning distinct years of publication, reveals a captivating panorama of the field's evolution and progression. In the last ten years, there have been significant advancements in this intriguing field of scientific investigation, revealing itself as an ever-expanding frontier of knowledge. The histogram chart effectively conveys the visual depiction, serving as compelling evidence of the dynamic nature of this journey. In the vast field of quantum technology, the primary focus of study and innovation primarily centers around the intriguing domain of QMs. This explanation demonstrates the changing field of quantum technology, highlighted by the prominent importance of QMs in current research efforts.

Over the past years, the Nobel Prize has been awarded in the field of the quantum domain, signing a rapid and transformative evolution in this domain. QMs play a crucial role in driving advancements in quantum technology, highlighting their significance in shaping the future of this field. QMs have gained substantial attention in materials science, yet the literature lacks comprehensive reviews due to their recent emergence. Addressing this gap, our paper focuses on introducing QMs, distinctive properties, and utilizing QMs in different applications. To summarize the whole quantum material domain in one paper is very difficult. We organize the paper in this way. It delves into the significant contributions made by the subject in various research areas, spanning a wide range of domains. It ultimately leads to a comprehensive discussion encompassing conclusions, prospects, and the challenges faced within QMs. It is a lengthy paper, but we try to summarize by figures, table, etc., and how the paper is organized is represented by Fig.2.This paper is a valuable resource for researchers, scientists, and engineers looking to better understand QMs and their applications.

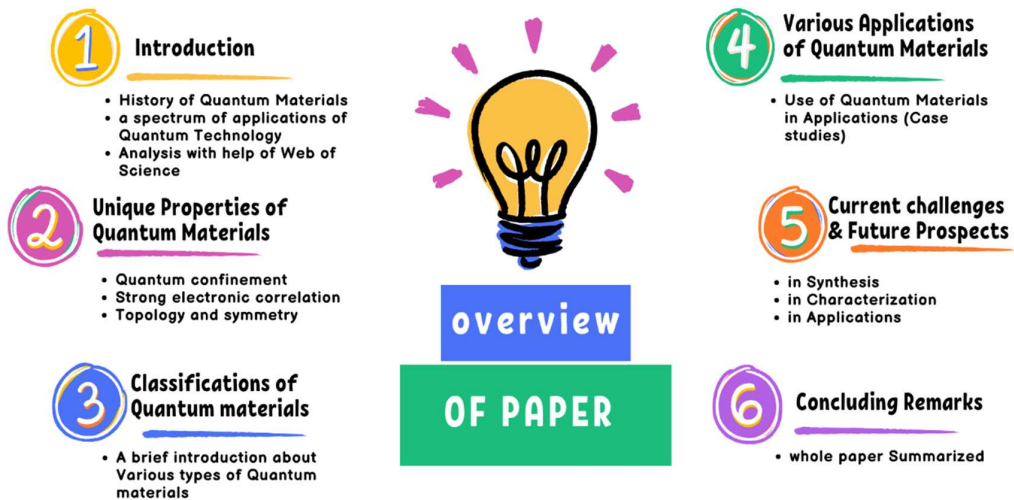

Fig. 2: a graphical representation of the article.

## 2. Basics of QMs and unique properties of QMs

In the introduction part, QMs are discussed very briefly about QMs. First, going ahead in this section, we discussed what QMs are and what a possible definition of QMs is. The term "Quantum Materials" is relatively new in the scientific community, leading to some confusion despite these materials being studied for a while. Research in this area is rapidly expanding, driven by the need for detailed consideration of electron wave functions and their surroundings. To Robert Cava, "I will not endeavor to provide a precise definition of what a quantum material is, but I know one when I see it." [20] Examples of QMs that you may generally be familiar with are those that display the quantum Hall effect or superconductivity, TIs, spin liquids, qubits, quantum sensors, or QDs. Interactions of fundamental degrees of freedom—lattice, charge, orbital, and spin—result in complex electronic states on the atomic scale.

Classical solid-state materials exhibit properties that can be primarily explained using classical particle behavior or a fundamental quantum mechanical treatment of electrons. For example, ionic motion in solids can be understood through electrostatic-based models, although a complete comprehension may require a quantum description of electronic states. Similarly, hybrid perovskites, like conventional semiconductors, may only demand physics within electron wave functions, albeit it can be insightful. In contrast, QMs demonstrate more esoteric yet tangible quantum effects such as fluctuations, entanglement, coherence, and dependence on the topology of quantum mechanical wave functions. TIs exemplify this, where surface electrons exhibit metallic properties distinct from bulk electrons, offering potential for energy-efficient electronics with resilient charge transport capabilities. This section will briefly discuss the fundamental properties of QMs —quantum confinement, strong correlations, and topological states.

## 2.1. Quantum confinement

It is a fundamental phenomenon in nanoscale materials, particularly semiconductors and other confined systems. It refers to the effect of restricting the motion of electrons and other quantum particles within a region smaller than their characteristic wavelength. This confinement leads to the quantization of energy levels, resulting in discrete energy states rather than the continuous energy bands seen in bulk materials.[37] Quantum confinement plays a crucial role in determining nanoscale materials' electronic, optical, and physical properties. Its unique nature explains why electrons in nanomaterials have higher energy than electrons in bulk materials. Depending on the size of the QD, confined electrons have more energy than those in bulk materials. When the dimensions of the semiconductor nanomaterials are shrunk from 2D to 1D or from 1D to 0D, they show exciting behavior. The quantum confinement effect may be present when the nanomaterials are smaller and smaller, i.e., less than 100–10 nm or even less. This is because of the separate set of electron energy levels, which cause changes in size (Figures 3 (a) and (b)). QDs make light in different colors depending on their size, which has led to their use in modern displays, biological imaging, and quantum cryptography.[38] To understand quantum confinement better, let us delve into its key aspects:

"Quantum confinement" is mainly used to discuss the energy of confined electrons (or electron holes). Compared with bulk solid materials, energy levels of the electrons of nanocrystals will not be the same as bulk. When you get the limited electron wave functions, they turn into a clear set of energy levels, as shown in Figure 3. These effects happen when the potential gets close to the de Broglie wavelength of electrons. This causes changes or discrete amounts of energy. The effects are called quantum confinement, and nanocrystals that have them are often called QDs. [39]

To learn more about quantum confinement, we need to know about QDs. It is now possible to see quantum confinement effects in a new type of material called QDs. QDs are nm-sized semiconductor crystals, and molecules comprise tightly packed electrons or pairs of electrons and holes called "excitons." When an electron is excited into a higher energy state, either through absorption of a photon or by another excitation method such as electroluminescence) This creates a positively charged space at the lower energy level, known as a hole. This results in the formation of the electron-hole pair. These two particles sometimes exist in a bound state, forming a single quasi-particle known as an "exciton." Within an exciton, the electron and the hole pair are bound together by coulombic interaction, and this strength of the bond is quantified by its exciton binding energy.

Excitons have an average physical separation between the electron and the hole pair, referred to as the exciton Bohr radius. Every semiconductor material has a characteristic exciton Bohr radius; depending on this, we classify them into Frenkel and Wannier excitons. Frenkel: Have a tightly bound radius of a similar magnitude as the unit cell, and Wannier: Have a large radius exceeding the size of the unit cell. While the de Broglie wavelength is very small compared to the size of the limiting structure, the particle acts like it is free. The energy states stay the same at this stage, and the bandgap returns to its original place. Another energy spectrum does not stay continuous; it breaks up into discrete waves as the size of the confining object shrinks to the nanoscale, as represented in Figure 3. So, the bandgap has properties that rely on particle size, which eventually leads to a blue shift in the light emitted as the particle size decreases. However, this result shows what happens when you keep the electrons and electron-hole pair (also called excitons) in a space close to the critical quantum limit, known as the Bohr exciton radius. The band gap ($E_g$) increases in magnitude (increase in energy of the band-to-band excitation peaks (blue shift)) as the semiconductor particle radius decreases in size to the point when it becomes comparable or smaller than that of the exciton radius as shown in Figure 3. This property has led to various fields: semiconducting quantum wells and superlattice devices, nonlinear optical materials, photocatalysis, and imaging systems.

Now, we will classify it into three regimes based on the exciton Bohr radius. Weak confinement regime: The radius of the crystallite is greater than the exciton Bohr radius. Moderate confinement regime: The radius of the crystallite is comparable to the exciton radius. Strong confinement regime: The radius of the crystallite is smaller than the Bohr exciton radius.[39] Also, depending on the confinement dimension, we can classify them as shown in Figure 3(b). Electrons confined in one direction, i.e., quantum wires: Electrons can quickly move in one dimension (1D), quantum wells (thin films): Electrons can promptly move in two dimensions (2D), so two dimensions are quantized.

Figure 3 (a) is a Representation of the confinement of QMs, their corresponding energy levels [40], and a schematic example of broken symmetry and the functional form of the density of states in one-dimensional (1D), two-dimensional (2D), and three-dimensional (3D) confined QMs. Where $E_{CB}$ energy of the conduction band, EVB is the energy of the valence band, ρ density of states, and n is no. of state per unit space. Figure 3 (b) demonstrates that when particle size decreases and reaches the nanoscale, it decreases in confining dimension and makes energy levels discrete, and this increases the band gap and also increases in band gap energy[40]

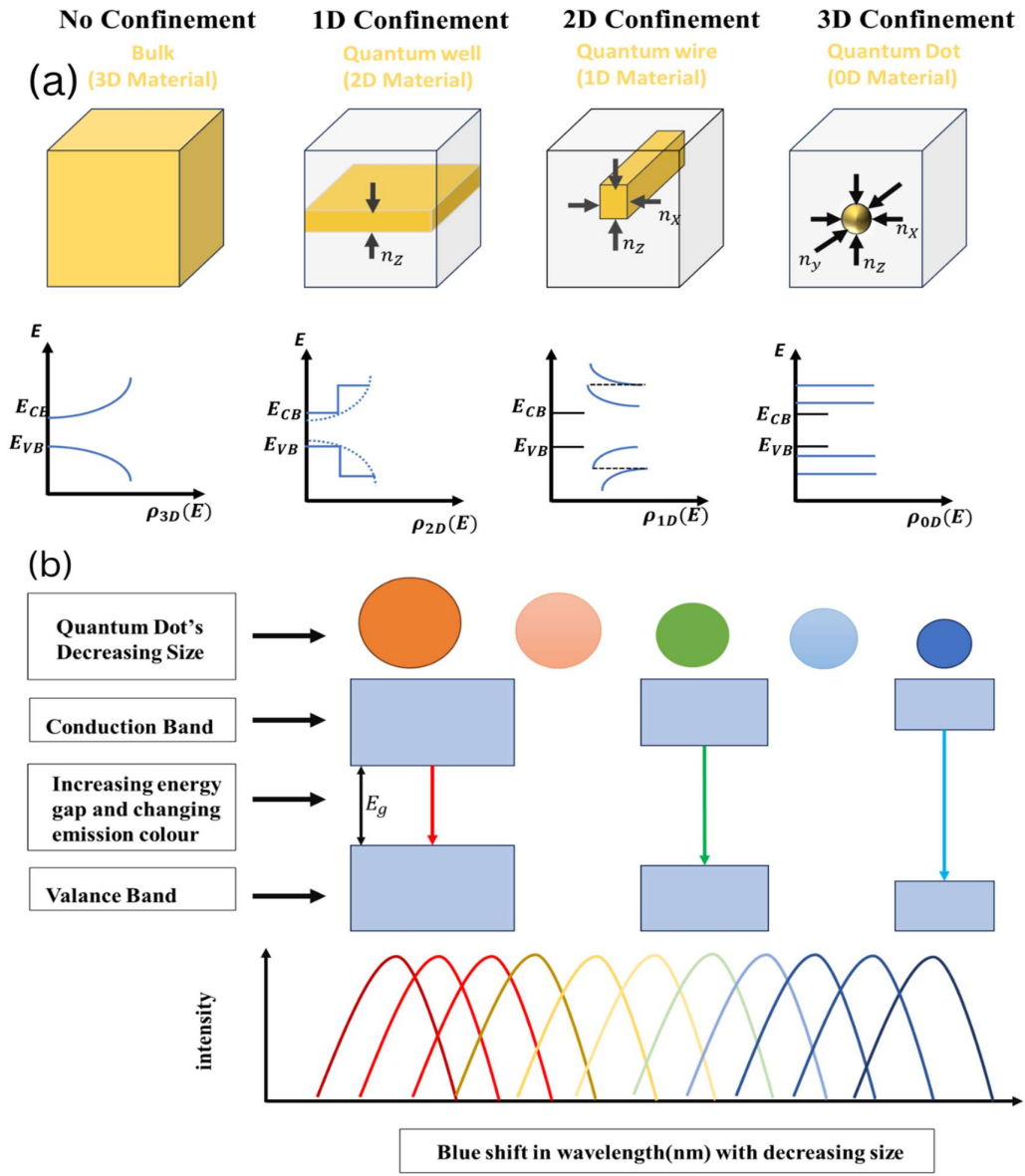

Figure 3: (a) Representation of confinement with corresponding energy levels and density of states of the QMs, (b) Bandgap energy variation with size of the QMs.

We use quantum mechanics to describe particles like electrons as discrete particles and waves with associated wavelengths. A particle's de Broglie wavelength (λ) is inversely proportional to its momentum (p). As the size of a confined region becomes comparable to or smaller than the de-Broglie wavelength of particles, their wave-like behavior becomes significant. In bulk materials, electrons have continuous energy bands due to the large number of atoms. However, in nanoscale materials, the confined dimensions impose boundary conditions that limit electrons' possible standing wave patterns (orbitals). This leads to the quantization of energy levels, where only specific energy values are allowed. The energy levels become discrete, resembling the rungs of a ladder rather than a continuous slope. The bandgap of a material is the energy difference between the highest energy electron in the valence band and the lowest energy electron in the conduction band. In bulk materials, the bandgap remains relatively constant. In nanostructures, especially semiconductor nanoparticles or QDs, the bandgap increases with decreasing particle size due to quantum confinement. This is pictorially shown in Figure 3. This means the electronic transitions responsible for optical properties occur at higher energies (shorter wavelengths) than in bulk materials.

Quantum confinement significantly affects the absorption and emission of light in nanoscale materials. As the energy levels are quantized, the energy difference between electronic transitions becomes more significant, resulting in more energetic absorption and emission peaks. This can lead to enhanced fluorescence, luminescence, and color-tunability of nanomaterials, which have applications in displays, lasers, and biological imaging.

In bulk materials, the conductivity is mainly determined by the concentration of free charge carriers (electrons or holes). In quantum-confined systems, the discrete energy levels create an energy gap between the highest occupied and lowest unoccupied states. This can hinder the movement of charge carriers and reduce the conductivity. However, quantum tunneling becomes more prominent at specific energy levels, allowing carriers to pass through energy barriers and leading to exciting conductivity behaviors. Quantum tunneling is a microscopic phenomenon where a particle can penetrate and, in most cases, pass through a potential barrier. The maximum height of the barrier is assumed to be higher than the particle's kinetic energy. Therefore, such a motion is not allowed by the laws of classical dynamics.[41]

Quantum confinement is crucial in the development of nanoscale electronic devices. QDs, for instance, can be used as single-electron transistors or quantum bits (qubits) in quantum computing due to their discrete energy levels and controllable charging behavior. In summary, quantum confinement is a phenomenon that arises in nanoscale materials where the wave-like nature of particles becomes pronounced due to spatial confinement. This confinement results in quantized energy levels, altered bandgap, enhanced optical properties, and unique electronic behaviour, all of which have wide-ranging applications in nanoelectronics, photonics, and materials science.

*2.2. Strong electronic correlation*

QMs exhibit unique electronic, magnetic, and optical properties from quantum mechanical effects at the atomic and subatomic levels. Strong correlation effects in QMs are a fascinating and intricate aspect that plays a crucial role in determining their behaviour. These correlation effects emerge due to the strong interactions between electrons within the material, leading to nontrivial behaviours that cannot be understood using classical physics alone. Electrons in classical materials frequently function as self-sufficient individuals, but in quantum QMs, significant interactions predominate due to factors such as Coulomb repulsion. This complex interaction results in emergent behaviours that deviate significantly from traditional classical expectations. High-temperature superconductivity, seen in certain QMs, is one notable example in which electron pairs termed Cooper pairs travel coherently along the lattice, resulting in zero electrical resistance. This phenomenon defies conventional wisdom and can transform power transmission and energy storage technology.[42] To understand strong correlation better, we will delve into the strong correlation properties of QMs and their significance in various physical phenomena.

In metals like copper and aluminum, mobile electrons interact slightly due to Coulomb forces but are largely unaffected by them due to their high kinetic energy. This allows for a perturbative approach based on single-electron theories. However, traditional theories struggle when Coulomb interactions become comparable to or exceed the kinetic energy, as in strongly correlated systems.[43] These materials offer fertile ground for discovering new physics, especially near zero-temperature phase transitions, where quantum critical fluctuations can lead to exotic excitations and novel quantum phases. Strongly correlated electron systems have driven significant theoretical advancements, including highly entangled phases of matter and quantum critical points that extend beyond conventional frameworks like Landau's theory.[44]

In QMs, electrons are the charge carriers, and the principles of quantum mechanics govern their behaviour.[45] When electrons interact strongly with each other, their collective behaviour can deviate significantly from that predicted by simple models. This interaction gives rise to various quantum phases, such as Mott insulators, correlated metals, and high-temperature superconductors. Many materials' electrical, magnetic, optical, and mechanical properties depend on electron-electron interactions. Hund's Rules in transition metals illustrate that Coulomb energy savings make d electron spin alignment and spatial orbital living profitable. Based on electron-electron repulsion and quantum physics, these and related exchange interactions underpin magnetic order, which is essential in industry and physics. The complicated interaction between electron-electron interactions, lattice structure, kinetic energy (via electron quantum-mechanical tunneling in crystals), and magnetic degrees of freedom is noteworthy. Ground states with differing symmetries and low energy excitations compete, as do "local" quantum degrees of freedom associated with lattice sites and longer-scale fluctuations and excitations. When an electronic population or magnetic field tilts the energy scale balance, new phases often develop at quantum phase transitions. These phases can have surprising and useful properties like high-temperature superconductivity. Technologically relevant features include orders-of-magnitude electrical conductivity shifts (metal-insulator transitions), perhaps modified by a magnetic field ("colossal magnetoresistance"). Strong correlations can function as a lever arm at certain behavior cusps, making modest controllable factor changes affect material attributes. Strong electron correlations enable complicated electrical material properties but make examination difficult. Well-developed perturbative approaches like expansion, local density Hartree-Fock approximations, Thomas-Fermi screening in metals, and others allow us to understand the electronic behaviour of relatively modest electron-electron repulsion compared to kinetic energy. Controlled theoretical techniques are limited when interactions are high, as in d- and f-electron systems. The problem is open except for 1D systems with analytical and computational methods. The "renormalization group" paradigm and Landau's adiabaticity principle save the day.

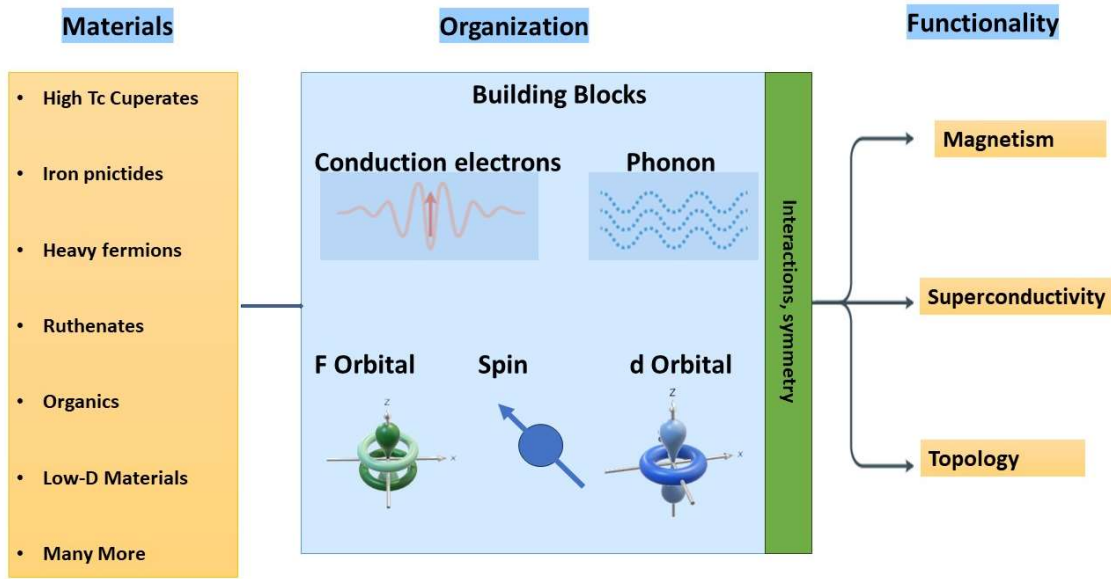

Figure 4: The operational characteristics of strongly correlated materials and their functionality with their organization.[46]

Figure 4 illustrates various materials housing strongly correlated electrons, including cuprate high-temperature superconductors,[47] ruthenates, transition metal oxides,[48] heavy fermion systems,[49] iron-based superconductors,[50] organics,[51] and low-dimensional materials.[52] These systems primarily involve d or f orbitals, which maintain some localization within the solid, resulting in heightened Coulomb interaction and narrower bandwidths compared to s or p orbitals. Furthermore, these orbitals generate localized moments within the materials under specific filling conditions. Figure 4 describes the functionality of strongly correlated materials by illustrating selected classes of such materials and how interactions between their low-energy degrees of freedom, often referred to as "building blocks" and symmetry, can give rise to various functionalities.

The Mott transition is a classic example of a strong correlation in materials. It is characterized by a phase transition from a metal to an insulator due to strong electron correlations. This transition is often associated with a marked change in the electronic structure, such as the transition from long to short c-axis layered perovskite structures. Factors like changes in orbital order drive the Mott transition and can occur together with structural transformations.[53] This phenomenon is a collective manifestation of the imbalance in the particle-wave duality of electrons. It is considered a quantum-critical transition in many materials, showcasing the intricate interplay between electron interactions and material properties.[54]

Another remarkable manifestation of a strong correlation is high-temperature superconductivity. These materials can conduct electricity without resistance at temperatures significantly higher than traditional superconductors.[55] In conventional superconductors, electron-phonon interactions pair electrons and allow them to flow without resistance. However, electron-electron solid interactions in high-temperature superconductors are thought to play a crucial role in enabling superconductivity at temperatures much higher than conventional theories predict.[56] Understanding and controlling these correlations are critical challenges in condensed matter physics. The discovery of high-temperature superconductors, such as copper oxide materials, has revolutionized the field of superconductivity, offering the potential for practical applications in various industries. High-temperature superconductivity challenges conventional theories of superconductivity and highlights the complex interplay of electron interactions in these materials, showcasing a strong correlation between electronic properties and superconducting behaviour.[57]

Quantum phase transitions occur at absolute zero temperature, representing a unique phenomenon in quantum systems.[58] These transitions are characterized by changes in the ground state of a system as specific parameters in its Hamiltonian are varied, leading to distinct phases at absolute zero temperature. Strong correlations can drive quantum phase transitions between ordered or disordered states. Unlike classical phase transitions driven by thermal fluctuations, quantum phase transitions at absolute zero are governed by quantum fluctuations and the interplay of quantum correlations. These transitions are exciting due to the novel and deep properties in the quantum critical region, offering insights into various behaviours in condensed-matter systems. The study of quantum phase transitions provides valuable information about the fundamental nature of quantum systems and their behaviour at the quantum critical point.

The lattice structure of a material indeed plays a crucial role in the emergence of strong correlation effects in quantum mechanical systems. The arrangement of atoms in a lattice influences the effective interactions between electrons, affecting their movement and interactions within the material. This connection between the electronic and lattice degrees of freedom is fundamental in understanding the behaviour of quantum mechanical systems, especially in strongly correlated electron systems where the interplay between electron-electron interactions and lattice dynamics is intricate. The lattice structure can modify the electronic properties of a material, leading to phenomena like high-temperature superconductivity or Mott transitions, highlighting the significance of considering both electronic and lattice aspects when studying strongly correlated quantum systems.[59]

Describing and predicting the behaviour of strongly correlated quantum mechanical systems pose a significant theoretical challenge due to the intricate interplay between different degrees of freedom. Traditional mean-field approximations often need to catch up in capturing the complexities of these systems. Advanced techniques such as dynamical mean-field theory (DMFT),[60] density matrix renormalization group (DMRG), and tensor network methods are employed to address these challenges. These methods offer more sophisticated approaches to model and understanding the behaviour of strongly correlated quantum systems by considering the dynamic nature of electron correlations and interactions. Techniques like DMFT focus on the local behaviour of electrons, providing insights into the effects of strong correlations on electronic properties. DMRG, on the other hand, is particularly useful for 1D systems, offering a powerful numerical method to study quantum many-body systems. Tensor network methods provide a framework to represent quantum states efficiently, enabling the study of entanglement and correlations in strongly correlated systems. By utilizing these advanced techniques, researchers can delve deeper into the complexities of strongly correlated quantum systems and gain a more comprehensive understanding of their behavior and properties.[61,62]

So far, we have discussed that the strong correlation properties of QMs are at the heart of many exotics and technologically relevant phenomena. Understanding and controlling these properties is key to developing novel materials with desired electronic, magnetic, and optical functionalities. QMs research provides insights into fundamental physics and paves the way for technological advancements in quantum computing, energy storage, and more.

## 2.3. Topology and symmetry in QMs

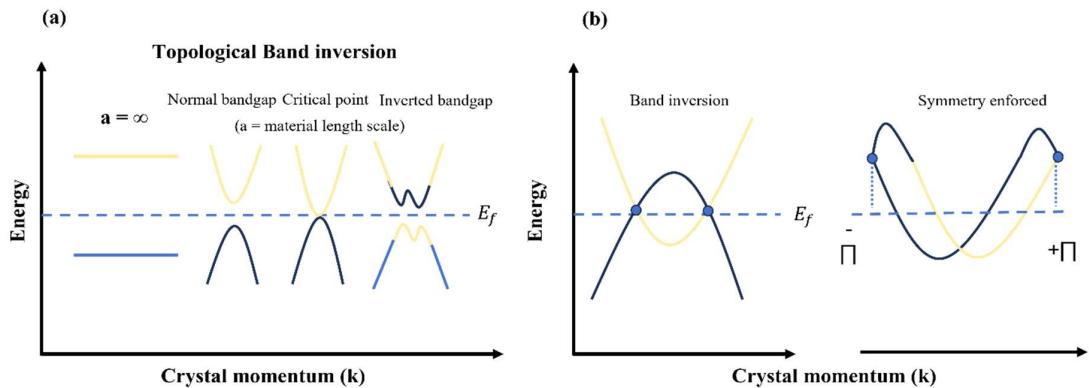

Figure 5: (a) Topology and band inversion in insulators, (b) Semimetal topological state classification

Figure 5 shows the topology and band inversion in insulators. a) Atomic limit energy levels determine the natural order of occupied (blue line) and unoccupied (yellow line) orbitals. As the lattice parameter (a) lowers, the gap between the occupied and unoccupied Bloch states closes at a critical point and reopens inverted concerning the atomic order, achieving a topological state. (b) Semimetal topological state classification. Materials can be bulk band inverted to create topological semimetals. Crystalline rotational or mirror symmetry protects the inverted valence and conduction band crossings (left). Dirac semimetals have four-fold degenerate crossing points shielded by rotational symmetry in the presence of inversion (P) and time-reversal (Θ) symmetries. Weyl semimetals have two-fold degenerate band crossings that form chiral Weyl fermions separated in momentum space. The distance between oppositely charged Weyl nodes protects the phase. Kramers–Weyl fermions and nonsymmorphic semimetals generate topological semimetals without bulk band inversion. Nonsymmorphic crystal symmetries protect spin-orbit coupling band crossings. Degeneracies and energy-momentum relationships near band touching points classify topological semimetals.

Condensed matter physics, focusing on particle systems' motion patterns and laws, has emerged as a vital subdiscipline. The discovery of the integer quantum Hall effect (IQHE) in 1980,[63] alongside subsequent findings of various topological quantum states, has unlocked a realm for global scientists. Topology and symmetry play crucial roles in QMs by influencing their properties and behaviour. In the realm of QMs, topology refers to the study of properties that remain unchanged under continuous deformations, like stretching or bending, without tearing or gluing. On the other hand, symmetry involves operations that leave a system invariant, such as spatial isotropy, translations, time reversal, and various internal quantum numbers. Topology and symmetry are fundamental concepts in QMs, influencing their phases, properties, and behaviour by defining symmetries, order parameters, and the structural characteristics that underpin their unique quantum phenomena. It is challenging to summarize everything here, but we will briefly discuss the main things here.

Topology is a mathematical idea that ignores small details and gradual changes. Think of a doughnut with a hole – the number of holes is a topological feature that tells us about its shape in space. This applies to materials, too, like how atoms are arranged in crystals or how electrons behave.[64] These special materials, called topological QMs, have unique qualities because of their topological structure. The discovery of the integer and fractional quantum Hall effects (IQHE and FQHE) has unveiled a unique form of matter known as the topological quantum state. Investigating the topological characteristics of these materials involves identifying similar properties in electronic band structures and physical quantities, leading to physical states less affected by material defects. [65,66]

The "topology of states" refers to how electron states are organized in a material, understood through unique numbers called "topological invariants." These numbers describe the shape of electron wavefunctions in momentum space.[67] A material with a "nontrivial topological number" has unique electron wavefunction shapes, guaranteeing special surface or edge states that are resilient to small changes in the material. An analogy is the unchanging basic shape of an object under gentle deformations. The appearance of these particular states in a material indicates a unique shape in its electron wavefunctions, as seen in the quantum Hall state (QHS), where the bulk becomes electrically insulating under a strong magnetic field in a 2D material. In contrast, chiral conducting states form along its edges. These edge states persist despite attempts to eliminate them, making QHS a precise standard for resistance and determining fundamental constants like the fine structure constant.

The band theory paradigm, a framework used in physics, has been used in predicting new topological materials and confirming their unique properties through experiments involving spectroscopy and transport. This analyzes band structures and topological invariants derived from topology and crystalline symmetry considerations. Recent advancements in band theory, which integrate topology and crystalline symmetries, have enabled a comprehensive classification of topological states. For instance, insulating nonmagnetic materials encompassing 230 space groups have been identified to support around 3000 different options.[64]

Topology serves as a keyway to organize and understand quantum states. It is all about the shape of the electronic wave function and how it behaves in k space, which is a mathematical representation of momentum space. The specific shape depends on the symmetry group of the crystal lattice. These topological characteristics control how materials respond to environmental changes and stay consistent even when the material's properties change smoothly. Imagine a scenario where the order of electron energy levels switches at specific points near the Fermi level. This change in order is like a band inversion, leading to a topologically nontrivial state.

Three main methods are commonly used to determine the unique properties of materials: 1) Directly calculating specific characteristics that indicate their topological nature and symmetry indicators.[68,69] 2) Employing a gradual transition to link the topological state of an unfamiliar material with that of a known one.[70] 3) Utilizing the relationship between bulk and surface states to identify the nontrivial properties of materials.[71] The first two methods involve analyzing the symmetry properties of bulk states, often represented in their irreducible forms at specific points in the Brillouin Zone. Initially conceived as a parity criterion for materials with inversion symmetry, this method was expanded to cover all types of space groups. The third method reveals the unique states of materials by examining their connections with bulk bands, offering experimental clues to confirm their distinctive properties.

Insulators are materials that do not conduct electricity, and they have a primary state where they're electrically inactive due to an energy gap between their filled and empty states. TIs are a newer insulator type with a gap in their bulk but can still conduct electricity along their surface because of unique metallic states caused by the arrangement of their bulk states. We can understand this by examining the material's structure with and without specific effects. If we remove a particular effect called spin-orbit coupling (SOC)[72,73], the material behaves like a regular insulator or a metal. But if we gradually introduce SOC, we observe a change from a regular state to a particular state where the gap in energy closes at a critical point, marking a transition. The nature of the material remains the same as long as its bulk properties stay intact. The surfaces of these materials act like a boundary between the unique interior and the regular outside, and this boundary results in unique surface states that connect the filled and empty states within the material. TIs are characterized by a property called $Z_2$, which can be either 0 or 1, indicating whether they are ordinary or extraordinary. This property can be calculated by analyzing the behavior of electronic states within the material. These unique properties also affect the behaviour of surface states, which have interesting features like linear energy patterns and a specific spin alignment that make them resistant to certain disruptions.

Topological states in semimetals are special because of their unique band structures close to the Fermi energy, which have distinct topological properties. These states are grouped based on their nodal band structures and surface states, with types such as Dirac, Weyl, or line-node semimetals identified by their node dispersion and the symmetries that protect these characteristics.[74,75] In semimetals, these topological band crossings can either happen accidentally, like Dirac points and lines, which are preserved by specific crystal symmetries and remain stable under small changes, or they can be enforced by symmetry, arising from nonsymmorphic symmetries and remaining stable even with significant deformations of the system. These topological states are crucial for various material properties and applications, showcasing unusual transport behaviours, durable surface states, and distinctive responses to electromagnetic fields. These topological semimetals, like Dirac and Weyl semimetals, have been supported by theoretical predictions from first-principles calculations [76], emphasizing the importance of understanding and exploring these materials for future technological progress.

One of the defining features of topological states is their remarkable robustness against local perturbations and disorder.[77] This robustness arises from the nontrivial topology of the electronic wavefunctions. The disorder can scatter electrons and degrade their behaviour in conventional materials, leading to effects like resistivity in electronic systems. As highlighted in the provided sources, this property is a pivotal advantage of topological systems, showcasing the absence of electronic backscattering in phenomena like the quantum Hall effect and spin-Hall effect. Topological states exhibit unidirectional waveguiding and edge states that persist despite imperfections or impurities, making them highly resilient to small perturbations and fabrication flaws. This robustness against disorder and defects allows for efficient energy and information routing in various 2D platforms, ranging from quantum electronics to classical photonic and phononic devices, offering significant application potential across different fields of research and technology.

In the study of TIs, specific key terms are vital to understand. TIs often have an "energy gap," a range of energy levels where electrons can't exist. This gap is crucial for stability, preventing unwanted energy fluctuations that could cause data loss.[78] Another important aspect is the emergence of "edge states" along boundaries like edges or surfaces. These states are localized and conduct electricity without scattering. This protection allows for robust information transfer and transport, even in impurities or defects.[79]

The Quantum Hall Effect (QHE) is a remarkable discovery in physics where, in 2D electronic systems, the Hall resistance is quantized in integer multiples of a fundamental constant.[80] This phenomenon arises from the unique topological characteristics of the system, resulting in the precise quantization of the Hall conductivity despite any disorder. Interestingly, this effect doesn't rely on electron interactions, as even a single particle in a magnetic field exhibits behavior similar to that of multiple particles. The quantization of the Hall resistance occurs in distinct plateaus, each corresponding to a specific number of filled Landau levels. Crucially, edge states along the boundaries of the system, known as chiral modes, play a pivotal role in maintaining the quantization of the Hall conductivity. These edge states, a consequence of the system's topological properties, remain robust against external perturbations. The QHE has revolutionized our understanding of condensed matter physics, inspiring new theories and driving experimental advancements, emphasizing the significance of topological properties in electronic systems under strong magnetic fields.[81,82]

Time-reversal symmetry plays a significant role in classifying topological states. Sometimes, TIs exhibit a $Z_2$ invariant [83], indicating whether the material is a trivial insulator or a nontrivial topological insulator. This $Z_2$ invariant is protected by time-reversal symmetry and determines the presence of protected edge states. TIs are materials with a bulk excitation gap generated by the spin-orbit interaction that differs from conventional insulators. This distinction is characterized by $Z_2$ topological invariants representing the ground state. A single $Z_2$ invariant distinguishes the ordinary insulator from the quantum spin-Hall phase in two dimensions. Four $Z_2$ invariants distinguish the ordinary insulator from "weak" and "strong" TIs in three dimensions. These phases are characterized by gapless surface (or edge) states.

The anti-de Sitter/conformal field theory (AdS/CFT)[84] correspondence is a theoretical framework that relates specific strongly interacting quantum field theories to higher-dimensional gravitational theories. Certain topological states, like TIs, can be connected to gravitational theories in a higher-dimensional space, suggesting a deep connection between topology and gravity.

In summary, topological states of matter are a remarkable manifestation of the deep connection between quantum physics and topology. These states exhibit robustness, energy gaps, edge states, and other distinctive properties that stem from their nontrivial topology. While much progress has been made in understanding and manipulating these states, they continue to be an active area of research with promising implications for both fundamental physics and technological applications.

Table 1: A collection of review papers that showcase the fundamental properties of QMs, along with brief descriptions of each paper.

| S.No. | Title of Paper | Description | Ref. |
|---|---|---|---|
| 1. | Emergent functions of quantum materials | QMs promise to revolutionize technology, from Mottronics to quantum computing, by harnessing the power of Majorana fermions to combat decoherence in solid-state qubits. Kitaev's 1D p-wave superconductor model presents intriguing possibilities for the future of quantum computing. | [85] |
| 2. | The physics of quantum materials | Spin-orbit coupling generates nontrivial electronic topology in materials like HgTe quantum wells, yielding TIs with edge states protected by time-reversal symmetry. These materials exhibit emergent quasiparticles with unique exchange statistics, showcasing novel phenomena such as dissipation less transport and exotic fractional quantum Hall effect states. | [10] |
| 3. | Towards properties on demand in quantum materials | Impulsive stimulation unlocks hidden quantum phases with potential for new time crystals, while Floquet engineering pioneers stable topological structures in cold atoms, inspiring novel strategies for condensed matter systems. Static perturbations in 2D materials and heterostructures enable precise control of material properties, including pseudomagnetic fields, opening doors to advanced electronic designs. | [10] |
| 4. | Topology & Symmetry in Quantum Materials | Topological materials herald a new frontier in fundamental science, offering remarkable electromagnetic responses rooted in crystalline geometry. With ongoing growth and unique opportunities, they promise to unravel the secrets of quantum matter and explore exciting new physics. | [86] |

| | | This abstract explores the intriguing world of QMs, where surface electrons exhibit unique properties, and quantum spin liquids give rise to exotic phenomena, promising energy-efficient electronics, and novel applications. Chemistry plays a pivotal role in understanding and harnessing these remarkable materials. | [20] |
|---|---|---|---|
| 5. | Introduction: Quantum Materials | | |

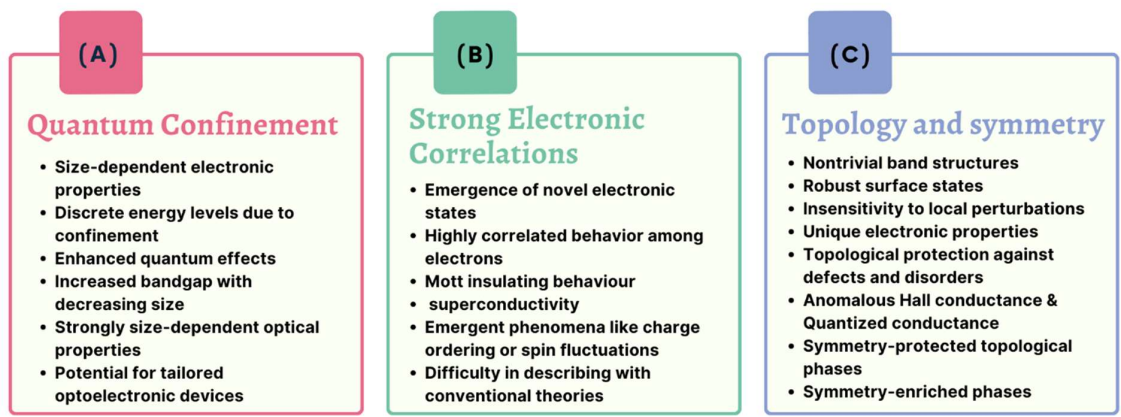

Figure 6: A pictorial representation of properties of QMs.

Summarizing the critical characteristics of QMs in a comprehensive review paper is a challenge. To facilitate comprehension, we have thoughtfully provided a helpful reference in Table 1, featuring citations to recent review papers that extensively explore the key properties of QMs. Figure 6 represents all the properties in one place. This table is a valuable resource, offering readers seeking a deeper understanding of the subject a compilation of review papers alongside succinct descriptions of their content. So far, in this section, we have summarized the essential properties and characteristics of QMs. With its essential qualities of quantum confinement, strong correlations, and topological states, QMs constitute a significant change in materials research. These features, which distinguish QMs from classical materials, enable QMs to serve as the foundation for breakthrough applications in various disciplines. We are getting closer to a future in which QMs transform industry, technology, and our knowledge of the fundamental nature of matter as we open the secrets of the quantum world.

## 3. Classifications of Quantum Materials

In the previous section, we discussed QMs and fundamental properties. Some materials that have been studied many times in past years and fit the criteria for QMs are also represented in Figure 7; we will discuss them in very brief one by one in this section.

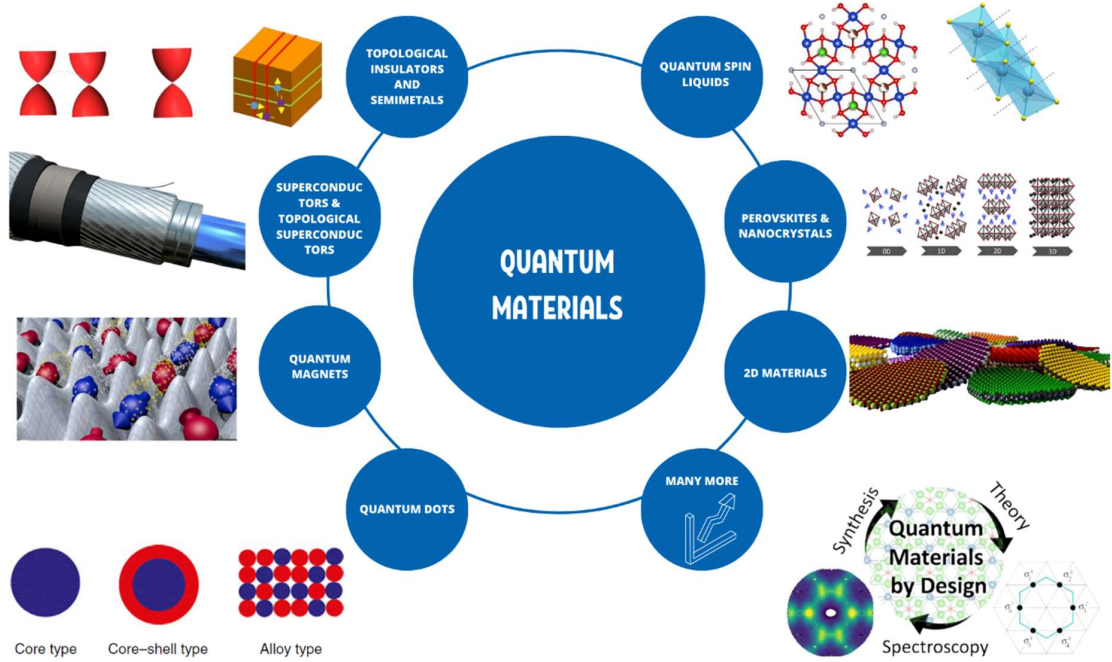

Figure 7: Exploring the diverse classifications of QMs for comprehensive understanding and further research.

## 3.1. Topological Insulators and Semimetals

Generally, solid materials are classified as semiconductors, insulators, and conductors. While TIs are a unique class of substances. A conductor has many no. of free electrons that move very fast. Insulators have a large bandgap, which restricts electron transport, while semiconductors have a moderate band gap between conductors and insulators. TIs do not fit under any of the categories stated. They function as surface conductors and core insulators in materials. We can imagine it like a block of wood wrapped in tin foil. Anirban wrote a very impressive review article in 2023 on behalf of 15 years of TIs.[87] Here, in short, we will summarize this.

Since the quantum Hall effect in a 2D electron gas system was found, QMs have gained much attention for having strange properties. TIs and semimetals have a reversed order of occupied valence bands and an unoccupied conduction band in the Brillouin zone. This is one way that they are similar to QMs. Dirac dispersion gives the surface of a TIs high mobility, but backscattering is impossible because of spin momentum locking. TIs show a 2D Dirac dispersion. Conventional insulators cannot support a chiral surface state, but these materials can accomplish this. Their characteristics are topologically protected, and because of their topological protection and immunity to moderate disorder, these traits generate a resistance that depends purely on fundamental physical constants. Although TIs have a bulk electronic state characterized by a small bandgap, which means that no free carriers are available in the bulk state, they possess a topologically protected state. The surface that resembles it. Because this surface has a Dirac point that can traverse the bandgap, the surface of TIs is composed of conductive material. Variations in TIs are caused by not all TI surfaces having the same number of Dirac points corresponding to them.[88]

Semimetals (also known as topological semimetals) are a different category of QMs distinguished by the absence of a gap between the valence and conduction states. Band, in addition to the fact that they are thought to have a gapless state and possess 3D Dirac dispersion. Semimetals can be divided into two categories: Dirac semimetals and Weyl semimetals. In the case of Dirac semimetals, a fourfold degenerate band is present. Weyl semimetals, on the other hand, have degenerate twofold points that are created by the linear crossing of two band points. These points are sometimes referred to as Weyl points. Therefore, they are characterized by obedience to inversion and time-reversal symmetries and thus require additional crystalline symmetries, such as rotational symmetry, to provide stability. The band inversion forms the topological Fermi-arc surface state at two paired Weyl nodes composed of quantum spin liquid.[88,89]

### 3.2. Superconductors

Superconductors, materials where quantum mechanics dictates their properties, exhibit zero electrical resistance at extremely low temperatures, a phenomenon known as superconductivity. Upon reaching a critical temperature, typically lower than traditional conductors, superconductors lose all resistance abruptly. This contrasts with conventional conductors, where resistance gradually decreases as temperature drops. This property is crucial for creating efficient and lossless quantum circuits. A loop of superconducting wire can sustain an electric current indefinitely. Dutch scientist Heike Kamerlingh Onnes discovered mercury's superconductivity at 4.2 K in 1911, marking the beginning of superconductivity research.[90] The Meissner effect, observed in superconductors, expels magnetic fields when cooled below their critical temperature, causing nearby magnets to repel.[91]

High-temperature superconductors, discovered more recently, offer practical advantages over traditional superconductors by requiring less extreme cooling. These materials have led to advancements in various technologies, including ultra-efficient electricity grids, fault current limiters, and particle detectors. Scientists aim to develop room-temperature superconductors, eliminating costly cooling methods and revolutionizing multiple industries with accessible, cost-effective superconducting solutions.

### 3.3. Topological Superconductors

Topological superconductors are a fascinating class of materials that exhibit unique properties due to their topological nature. These materials are distinct from conventional superconductors and are characterized by Majorana fermions, exotic particles that can exist as their antiparticles.[92] The relationship between topological superconductivity and Majorana fermions is a key focus in these materials, with researchers exploring the fundamental concepts, basic theories, and expected properties of topological superconductors.[93]
The realization of topological superconductors involves intricate routes often relying on spin-orbit coupling and specific pairing symmetries. The quest for intrinsic topological superconductors has led to significant advancements in understanding the role of symmetry, defect states, and boundary effects in these materials. By elucidating the theory behind topological superconductors and exploring various materials realizations, scientists aim to unlock the full potential of these exotic states of matter for future technological applications.[94]

### *3.4. Quantum Magnets*

Quantum magnets have a rich history that dates back to ancient times. The discovery of magnetism is attributed to a shepherd named Magnes in Magnesia, Greece, who observed the attractive properties of magnetite when his iron stick and nails were drawn to a magnetic rock. Over the centuries, magnets have played crucial roles in various applications, from aiding navigation with compasses to modern uses like data storage in computers and generating electricity in generators. The understanding of magnets evolved significantly with the discovery of electromagnetism by Hans Christian Orsted in 1820, linking electricity and magnetism. This laid the foundation for the exploration of quantum aspects of magnetism.[95]

In contemporary research, quantum magnets are at the forefront of scientific inquiry, delving into complex phenomena like frustrated quantum magnetism and topological states of matter.[96] Studies explore emergent properties in materials where quantum effects dominate, leading to novel phases of matter like quantum spin liquids. Researchers employ advanced techniques like exact diagonalization to study large-scale systems and investigate phenomena such as chiral spin liquids in frustrated magnetic structures. These investigations deepen our understanding of fundamental physics and promise future technological advancements harnessing the unique properties of quantum magnets.[97]

### *3.5. Quantum spin liquids*

Magnetism is a crucial property of materials with several uses in electronics and energy storage. Quantum mechanics explains magnetism. Magnetism is caused by electric charge movement. When brought close together, magnets repel or attract each other based on their nature. Like electric forces, magnetic forces exhibit both attraction and repulsion. Two magnetic poles always exist. A magnet can be divided in half, with each half retaining its north and south poles. Electrons rotate around the nucleus and its axis. Spin creates a magnetic dipole, making electrons rotate around their axis like tiny magnets.[98,99] Materials contain many electrons, each generating a small magnetic field and interacting with the other electrons to form a collective stabilized magnetic field of a material.[100]

In 1973, Phil Anderson was the first to propose the quantum spin liquid state as the ground state for a spins system on a triangular lattice exhibiting antiferromagnetic interactions with their nearest neighbours.[101] The frustrated-magnets model can be utilized to describe a quantum spin liquid. This type of liquid is characterized by the arrangement of electron spin, which precludes an ordered alignment and results in a state that acts like a fluctuating liquid. At temperatures below zero, when the spin of conventional magnets is locked, the electron spin in a quantum spin liquid never aligns and instead fluctuates continually. One of the most renowned examples of a quantum spin liquid is Herbertsmithite, a copper-based mineral with the chemical formula $ZnCu_3(OH)_6C_{l2}$. Although only some examples of quantum spin liquids have been generated synthetically, scientists are doing their best to develop them. In 2003, Shimizu et al. presented the gapless spin phenomena in [κ-(BEDT-TTF)$_2$Cu$_2$(CN)$_3$].[102]

*3.6. Perovskites*

The first perovskite phase was considered a substance with the same crystal structure as the mineral calcium titanium oxide. Most of the time, perovskites comprise the chemical formula ABX$_3$. A and B are cations, and X is an anion linking to A and B. Perovskite shapes can be made by combining different elements. Perovskite materials have emerged as promising candidates for QMs due to their unique properties and potential applications in various fields. These materials exhibit high efficiency, low processing costs, and tuneable characteristics, making them attractive for photovoltaics and beyond. The synthesis of perovskite-inspired materials, including hybrid halide double perovskites, presents challenges such as high-temperature requirements and complex fabrication processes. Despite these challenges, ongoing research efforts are focused on enhancing perovskite materials' stability, efficiency, and environmental friendliness, aiming to develop stable, low-cost, and non-toxic alternatives for highly efficient multijunction solar cells.[103]

The field of perovskite materials is advancing rapidly. Exploring new compositions and structures, such as quasi-2D perovskite materials and lead-free alternatives, is expanding the compositional space of perovskites.[104] By incorporating larger organic cations and exploring novel synthesis routes, researchers are pushing the boundaries of perovskite materials, aiming to maximize their properties and performance. The integration of 2D materials with perovskite solar panels and the development of large-scale perovskite modules are just a few examples of ongoing advancements in the field, highlighting the significant potential of perovskite materials in quantum applications and energy-relevant technologies.[105]

*3.7. 2D materials*

2D materials have sparked immense interest in the realm of QMs due to their remarkable properties at the nanoscale. These materials, composed of single or few atomic layers, exhibit exotic quantum phenomena that defy classical understanding. 2D materials have revolutionized the materials science landscape since the discovery of graphene in 2004. Graphene, a single layer of carbon atoms arranged in a hexagonal lattice, showcased exceptional properties like high electrical conductivity, mechanical strength, and thermal conductivity, sparking a wave of research into other 2D materials.[106] Transition metal dichalcogenides (TMDs), phosphorene, and hexagonal boron nitride (hBN) are among the diverse range of 2D materials explored for their unique characteristics arising from reduced dimensionality, enabling the study of quantum confinement effects and emergent phenomena. Their ultrathin nature enables quantum confinement effects to dominate, leading to unique electronic, optical, and mechanical properties. Graphene, the most famous among them, boasts exceptional electron mobility, paving the way for novel applications in electronics and photonics. Moreover, 2D materials like transition metal dichalcogenides (TMDs) exhibit strong spin-orbit coupling and valley physics, opening avenues for quantum computing and spintronics.[107]

The discovery of the Quantum Hall Effect (QHE) within 2D electron systems has revolutionized our understanding of condensed matter physics.[108] This interesting phenomenon, where the Hall resistance demonstrates quantized values, illustrates the intricate interplay between electron-electron interactions and topological properties. Whether integer or fractional, the QHE has paved the way for exploring emergent quantum phenomena, such as the Quantum Spin Hall Effect and the Quantum Anomalous Hall Effect. Research on Quantum Hall Systems scrutinizes edge and bulk properties, unraveling insights into electron behavior in confined geometries. This theoretical framework advances fundamental physics and holds promise for applications in quantum computing and TIs.

QMs research in the context of 2D materials delves into harnessing and controlling their quantum properties for technological advancements. By engineering heterostructures comprised of different 2D materials or integrating them with other materials, researchers aim to tailor their electronic band structures and create designer quantum systems. This pursuit has led to discovering phenomena such as moiré superlattices in twisted bilayer graphene, offering a platform for studying correlated electron physics and unconventional superconductivity. Furthermore, the ability to tune the electronic and optical properties of 2D materials through external stimuli like strain or electric fields holds promise for developing next-generation quantum devices with enhanced functionality.

### 3.8. Quantum Dots a specific type of nanomaterial

On the other hand, nanomaterials refer to materials with dimensions on the nm scale in at least one dimension. This broad category includes many materials, such as nanoparticles, nanowires, nanotubes, and more. Nanomaterials can be made from various substances, including metals, semiconductors, polymers, and ceramics. They exhibit unique properties compared to their bulk counterparts due to their small size, high surface area-to-volume ratio, and quantum effects. QDs are tiny semiconductor particles with dimensions typically ranging from a few nm to a few hundred nm. At this scale, they exhibit quantum mechanical properties, such as discrete energy levels, due to their small size. While QDs are a specific type of nanomaterial, not all nanomaterials are QDs. QDs are semiconductor nanoparticles with distinct quantum properties, while nanomaterials encompass a broader range of materials with dimensions on the nm scale. Because QDs are so small, quantum mechanics can accurately describe their properties. QDs are zero-dimensional materials with diameters of less than 10 nm. Size, composition, and structure are the three factors that have an impact on the properties of QDs, and there are three different types of structure: core type, core-shell type, and alloy type, also mentioned in Figure 7. The rare earth metals, actinides, alkaline earth metals, and other p-block elements (groups 13–16) are commonly used in synthesizing QMs.

Core-type QDs are solely made up of a core (as shown in Figure 7) and can be made out of single-component materials with uniform internal dimensions. Composition primarily comprises elements belonging to groups II–VI, III–V, or IV. CdSe [109], CdS [110], and CdTe [111] are the most well-known cadmium-based QDs in this category because of their visible-range photoluminescence (PL). These QDs are employed in several applications and are prevalent in this area. On the other hand, Cadmium's toxicity restricts its use in certain situations. There has been a shift in emphasis towards QDs that do not contain heavy metals, such as group IV QDs, which include carbon CQDs. There are several benefits associated with CQDs, including low toxicity[112], high water solubility [113], electrochemical luminescence[114], biocompatibility[115], and emission wavelengths in the UV-Vis range.[116]

In QDs, PL is induced by electron–hole pair recombination.[117] This recombination can also occur through processes that do not involve radiation, decreasing PL's quantum yield.[118] To create core–shell QDs, higher-band materials used as shells are grown atop core materials. The quantum yield of QDs can be improved by coating them with shells that passivate nonradiative recombination sites. This also makes the QDs more resistant to the processing conditions used in various applications.[119] For instance, CdSe–ZnS core–shell QDs have higher than 50% quantum yields.[110]

Alloyed QDs comprise two or more distinct materials or components, each of which contributes to the attributes of the QDs while maintaining the same size throughout the alloy. Alloyed QDs containing multiple semiconductors exhibit mixed or intermediate optoelectronic qualities. This is because these semiconductor nanostructures give different and nonlinear optoelectronic properties. Compared to simple CdSe QDs, the quantum yield of core–shell and alloyed $CdSe_{1-x}Te_x$ can be improved by as much as sixty percent, as stated in the reference.[120]

This section describes the various types of well-known QMs in the literature. Table 2 summarizes different types of QMs and their corresponding vital properties and multidisciplinary applications.

Table 2: Representation of various type of QMs and corresponding key properties and their multidisciplinary applications.

| S.No. | Quantum Material | Key Properties | Applications |
|---|---|---|---|
| 1. | Superconductors | Zero electric resistance, perfect diamagnetism | Quantum computing, energy transmission, medical imaging |
| 2. | Quantum Dots | Tenable energy levels, size-dependent properties | Optoelectronics, quantum cryptography, drug delivery |
| 3. | Topological insulators | Conducting surface states, insulating bulk | Spintronics, quantum computing, advanced electronics |
| 4. | Graphene | Exceptional electron mobility, 2D structure | Electronics, transparent conductors, sensors |
| 5. | Quantum wells | Quantum confinement effects | Lasers, photodetectors, solar cells |
| 6. | Superfluid Helium | Zero viscosity, Quantum vortices | Cryogenic applications, fundamental physics studies |

| 7. | Majorana Fermions | Non-Abelian anions, robustness against decoherence | Topological quantum computing fault-tolerant qubits |
| --- | --- | --- | --- |
| 8. | Nitrogen-Vacancy centres | Spin defects in Diamond, long coherence times | Quantum sensing, magnetic resonance imaging, quantum information processing |
| 9. | Quantum Hall Effect | Discrete quantization of Hall conductance | Metrology, fundamental constant determination |
| 10. | 2D Materials | High carrier mobilities, superconductivity, good thermal conductivity | Electronics, Photonics, Sensing devices |

## 4. Applications of QMs

QMs, at the forefront of materials science, have ignited a revolution in technology and applications. These remarkable materials, with their unique electronic and optical properties arising from quantum mechanical effects, hold immense promise for a wide range of exceptional applications. From ultra-efficient photodetectors to quantum computing, QMs drive innovation and redefine the boundaries of what is possible in electronics, energy storage, and beyond. As we delve deeper into their properties and harness their potential, we are poised to unlock a new era of technological advancement that will shape the future in ways we are only beginning to imagine. In the following subsections, we will provide significant experimental studies reported in the last five years of QMs from the perspective of potential applications of QMs. We categorize applications into subsections, as represented in Figure 8.

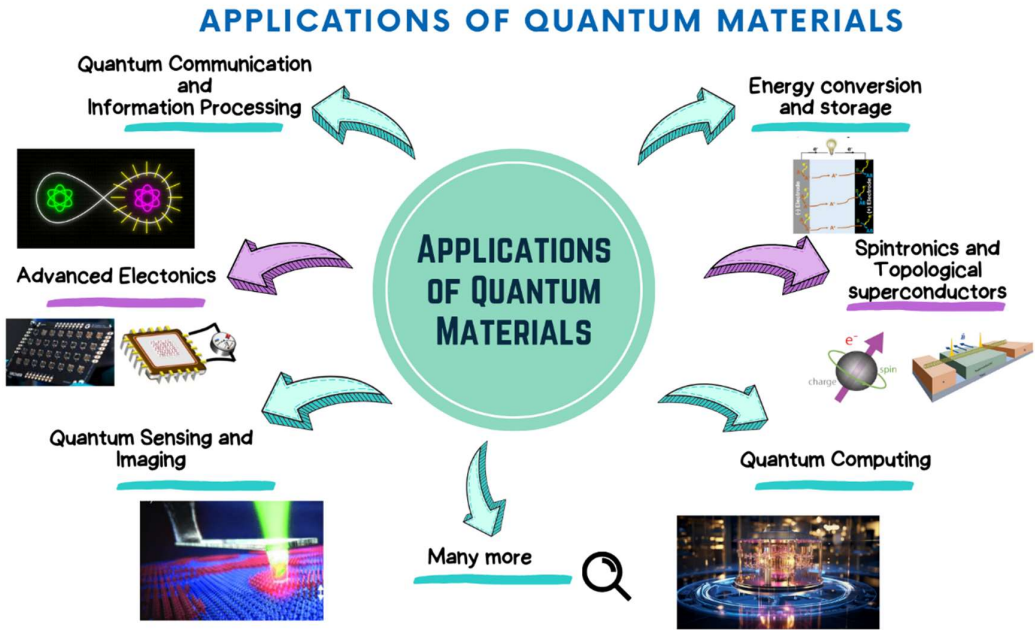

**Figure 8:** In brief wide range of QMs applications, promising innovative solutions and advancements across various fields.

*4.1. Advanced Electronics*

The emergence of quantum material-based electronic devices represents an innovative shift in the realm of technology and materials science. QDs, nanowires, and 2D materials have garnered significant attention for their remarkable properties and potential applications. QDs, nano-sized semiconductor particles, exhibit unique quantum confinement effects that enable precise control over their electronic and optical properties.

On the other hand, nanowires offer a platform for constructing ultra-compact, high-performance devices due to their exceptional electronic transport properties and versatile integration possibilities. 2D materials, such as graphene and transition metal dichalcogenides, present an exciting avenue for developing ultra-thin and flexible devices with exceptional electrical, thermal, and mechanical characteristics. These quantum material-based electronic devices hold promise across various fields, including electronics, photonics, energy harvesting, and quantum computing, fuelling a new era of innovation and pushing the boundaries of what is achievable in modern technology. This section will discuss the emergence of QMs-based electronic devices, including QDs, nanowires, and 2D materials, and explore their potential in next-generation transistors, sensors, and high-performance computing. These materials harness the unique properties of quantum mechanics to create devices with superior performance characteristics compared to traditional silicon-based electronics.

In 2017, Takeda Hirotaka introduced an innovative approach to fabricating nanowires to create a remarkably transparent and electrically conductive nanowire film.[121] The proposed method involves the synthesis of nanowires featuring a linearly interconnected structure formed by multiple particles. Notably, the nanowires are engineered with a carefully controlled diameter ratio falling within the range of 15 to 25 (A/B), ensuring that the resulting nanowire film exhibits outstanding transparency and excellent electrical conductivity. This breakthrough holds significant promise for applications requiring materials with dual transparency and electrical conduction characteristics, opening new avenues for advanced technologies in fields such as electronics, optoelectronics, and beyond.

In 2020, Jianjun Zhong successfully synthesized silver nanowires (AgNWs) through a polyol reduction process, employing $FeCl_3$ and NaBr as reaction inhibitors.[122] Notably, they found that varying the molar ratio of $FeCl_3$ to NaBr influenced the yield of AgNWs but had minimal impact on their morphology. These AgNWs were then utilized to fabricate transparent conductive films on glass slides, exhibiting commendable electrical conductivity, high transmittance, and a smooth surface texture. The AgNWs film heaters displayed rapid heat response characteristics and efficient electro-thermal conversion, achieved with low input voltages. Furthermore, they observed that these AgNWs films remained structurally intact up to 170°C, beyond which they fused into discontinuous segments at 200°C. Overall, his findings mark the promising potential of AgNWs for transparent and efficient heating applications, while shedding light on the nuanced relationship between reactant ratios and nanowire yield.

In 2022, Qikun Xu developed a novel gas sensor that uses black phosphorus QDs (BP QDs) modified $Ti_3C_2T_x$ nanosheets ($BQ/Ti_3C_2T_x$) as its sensing channel. This sensor shows remarkable performance in $NO_2$ detection at room temperature.[123] This sensor boasts a wide linear detection range, rapid response time, and exceptional specificity. Furthermore, the study delves into the influence of humidity and temperature on gas detection and introduces a practical calibration approach to mitigate humidity's impact, thereby ensuring the precision of $NO_2$ detection. The sensor's heightened sensitivity and selectivity can be attributed to the augmented adsorption energy of $BQ/Ti_3C_2T_x$ towards target gas molecules and the formation of P-O-Ti bonds. This research provides valuable insights into developing sensitive and selective gas sensors, highlighting the promise of functionalized MXene materials in this field.

A recent study by Yanzhe Li in 2023 showcases the promising application of transfer printing QDs as an effective colour-converting layer for advancing full-colour micro-LED displays.[124] Through the strategic design and creation of a distributed Bragg reflector (DBR), the study successfully enhances the colour purity of red and green light emitted by QDs while optimizing the utilization of blue light from micro-LEDs. This technology's implementation involves transferring red QDs onto the DBR as a colour-converting layer, subsequently integrated with blue micro-LEDs. This innovative approach facilitates the down-conversion of blue emission from the micro-LEDs into vibrant red emission, thereby realizing a full-color display. This research presents a significant step towards enhancing the quality and versatility of micro-LED displays, with potential applications ranging from consumer electronics to advanced visualization systems.

Mahima Chaudhary, in 2023, explored the utilization of nitrogen-doped carbon QDs (CQDs) on graphene to develop field-effect transistor-based (FET-based) optoelectronic memories that are programmable via UV illumination.[125] The study showcases the successful creation of such memories, emphasizing the pivotal role played by the charge-trapping properties of nitrogen-doped CQDs in ensuring long-lasting memory retention. Additionally, the interaction between the CQDs and graphene FET enables a photo-gating effect, enhancing the versatility of these memories. Notably, the research reveals that memory erasure can be achieved by applying a positive gate bias, facilitating the removal of trapped charges through recombination. This work marks the immense potential for engineering high-performance all-carbon, non-volatile FET-based optoelectronic memories by manipulating and coupling charge-trapping properties between colloidal CQDs and graphene.

In 2023, Yuanqing Zhou researched metal halide perovskite QDs, which possess remarkable optical properties that make them promising candidates for LED devices.[126] However, the conventional ligands used in these QDs have limited their practical application due to their instability. In an exciting breakthrough, incorporating the dual-function ligand DDA-MeS has proven to be a game-changer. This innovative approach has significantly enhanced the electrical performance of green QDs-LED devices and yielded impressive results. With DDA-MeS, the external quantum efficiency soared to 10.18%, the maximum brightness reached an impressive 8025 cd/m$^2$, and the turn-on voltage was reduced to just 2.5 V. This success highlights the tremendous potential of dual-function ligands in advancing the performance of perovskite quantum dot LED devices, paving the way for brighter and more energy-efficient lighting solutions.

Jinguo Ca in 2023 has achieved a significant breakthrough in lighting technology.[127] They successfully synthesized multicolour nitrogen-doped graphene QDs (NGQDs) through a one-pot method and refined the process further with column chromatography purification. These NGQDs exhibited three distinct emission colours: vivid blue, captivating cyan, and radiant yellow. Leveraging this innovation, the team integrated these NGQDs with InGaN chips to engineer LEDs capable of emitting these specific colors. Most notably, by strategically reducing the number of yellow NGQDs used in the blend, they created a white LED (WLED) with precise color coordinates of (0.324 and 0.334). This achievement paves the way for developing highly customizable and energy-efficient lighting solutions with broad applications in various fields.

Fidaa S.M. Ali, In 2022, found an intriguing phenomenon of the metal-insulator transition (MIT) within WSe$_2$ devices, focusing on the pivotal role played by the thickness of the WSe$_2$ channel and the associated unscreened charge impurity and trap density near the interface.[128] Our investigation revealed a compelling relationship between the thickness of the WSe$_2$ channel and the behaviour of the MIT. Specifically, as the WSe$_2$ channel thickness decreases, we observed a pronounced increase in critical carrier and trap densities. This effect engendered significant potential fluctuations and charge density heterogeneity within the system. Notably, this tuneable MIT behaviour is predominantly attributed to the substantial rise in trap density as WSe$_2$ thickness diminishes. Remarkably, our findings highlight a remarkable congruence between thinner WSe$_2$ devices and percolation theory, starkly contrasting the behavior exhibited by multilayer WSe$_2$ devices, which defy percolation theory. This study marks the critical importance of thickness control in manipulating the metal-insulator transition dynamics in WSe$_2$ devices and paves the way for novel applications in the field of nanoelectronics.

In 2022, Guen Hyung Oh showcased 2D transition-metal dichalcogenide (TMD) semiconductors, namely $MoS_2$, $MoTe_2$, and $WSe_2$, which were skillfully employed to construct van der Waals vertical heterostructures for tunneling field-effect transistor (TFET) applications.[129] The fabricated TFETs featuring $MoS_2$/$MoTe_2$ or $MoS_2$/$WSe_2$ heterostructures exhibited remarkably low subthreshold swings of 9.1 mV/dec and 7.5 mV/dec, respectively. This remarkable achievement signifies the efficient band-to-band tunneling processes harnessed in these devices. The electrical characteristics of these heterostructures were thoroughly investigated, revealing negative differential transconductance, negative differential resistance, and temperature-dependent I-V characteristics, all of which solidified the band-to-band tunneling mechanism. Incorporating atomically thin 2D materials and van der Waals vertical heterostructures in TFETs showcases substantial potential for advancing the development of next-generation wearable, stretchable, and flexible low-power electronic devices.

An impactful study In the domain of biosensing by Silvia Rizzato emerged field-effect transistors (FETs) as indispensable tools, boasting an array of advantages including exceptional sensitivity, rapid response times, the potential for miniaturization, and the convenience of portable read-out systems.[130] The integration of 2D nanomaterials into FETs has yielded auspicious outcomes, showcasing superior performance characteristics such as an impressive surface-to-volume ratio, enhanced carrier mobility, localized gating capabilities, heightened transconductance, and the ability to operate efficiently at low voltages. This paper delves into a comprehensive review of the progress and prospects of 2D materials in FET biosensors, shedding light on the myriad opportunities, recent applications, associated challenges, and forthcoming possibilities. Indeed, integrating 2D nanomaterials into biosensors is key to unlocking heightened sensing capabilities and optimizing device performance. With their compact size and remarkable sensitivity, FETs are poised to revolutionize real-time monitoring across a broad spectrum of biomedical diagnostic applications.

A recent study in electronics materials by Taoyu Zou [131] introduced a Solution-processed 2D transition metal dichalcogenides (TMDs) hold significant promise for the future of Complementary Metal-Oxide-Semiconductor (CMOS) technology, thanks to their atomic thinness and impressive electrical and mechanical properties. Nonetheless, the journey towards their practical application presents hurdles. Achieving scalable, high-purity 2D semiconductor mono- and few layers with sizable lateral dimensions and a narrow thickness distribution remains a formidable challenge. Furthermore, the field-effect mobility of solution-processed 2D TMD transistors lags behind counterparts produced through alternative methods, with the underexplored realm of solution-processed p-type 2D TMD transistors adding to the complexity. Despite these obstacles, there have been notable strides in developing solution-processed CMOS devices utilizing both n-type and p-type 2D TMD transistors bolstered by advancements in deposition techniques, device fabrication, and CMOS applications. As this field progresses, ongoing research endeavours will be essential to surmount current limitations and steer this technology toward its full potential.

This section explored QMs-based electronic devices, representing a significant leap forward in electronic engineering. Their unique properties enable precise control and manipulation of electrons, leading to improved device performance in transistors, sensors, and high-performance computing applications. Nowadays, various types of 2D materials are used in different electronics applications. As research continues in this field, we can expect to see increasingly innovative and powerful quantum material-based devices shaping the future of electronics.

*4.2. Quantum Sensing and Imaging*

QMs are revolutionizing sensor technology and imaging. They offer unique properties like superconductivity and magnetism, enabling ultra-precise sensors for medical imaging and environmental monitoring. Integrating QMs into techniques like MRI and microscopy enhances our ability to study atomic and molecular structures. Applications include highly sensitive magnetometers (SQUIDs) for medical diagnostics and quantum dot sensors for biomolecular sensing. Interferometry involving QMs, such as Bose-Einstein condensates and atom interferometers, offers unparalleled precision in measuring gravity and inertial forces. These sensors can detect minute changes in acceleration and rotation and find applications in navigation, geodesy, and tests of fundamental physics. In this section, QMs drive a new sensor development and imaging era, propelling scientific discovery to new heights.

A study By Alexander G. Milekhin in 2016 on surface-enhanced Raman scattering (SERS) within semiconductor nanostructures.[132] It investigates the manifestation of SERS through confined optical and surface optical phonons in a diverse range of materials, including CdS, CuS, GaN, and ZnO nanocrystals, along with GaN and ZnO nanorods and AlN nanowires. The findings reveal that SERS by phonons exhibits a resonant nature, with the most substantial enhancement observed in CdS nanocrystals and surface optical modes within ZnO nanocrystals. Moreover, this study unveils the augmentation of Raman scattering due to confined optical phonons and the emergence of novel Raman modes with distinct frequencies distinct from those found in ZnO bulk. These phenomena are attributed to surface optical modes within the nanostructures and are validated through calculations employing the dielectric continuum model. Furthermore, SERS by phonons is harnessed as a powerful tool for exploring the phonon spectrum in nanocrystal ensembles characterized by an exceptionally low areal density when coupled with metal plasmonic nanostructures. This research significantly advances our understanding of the intricate interplay between phonons and nanostructures. It offers valuable insights into their resonant behavior and the possibilities for phonon spectroscopy in the nanoscale regime.

Robin Corgier, in 2020, introduces a novel source engineering concept aimed at enhancing precision atom interferometry for binary quantum mixtures with extended drift times.[133] The authors created a multi-stage, non-linear atomic lens sequence to generate dual ensembles characterized by ultra-slow kinetic expansion energies, specifically below 15 pK. This innovation serves the crucial purpose of mitigating wavefront aberrations, a prominent source of systematic errors in precision atom interferometry. Furthermore, the authors developed a scaling approach to address the non-linear dynamics of binary quantum mixtures, thus expanding the potential applications of this technique in scenarios involving various intra- and inter-species interaction regimes. This research opens exciting possibilities for achieving more precise and stable atom interferometry measurements over extended timeframes.

Nitrogen-vacancy (NV) centers in diamonds are quantum defects with unique spin properties that allow them to sense magnetic and electric fields with high sensitivity. NV centers can detect small magnetic signals, such as in biomagnetism imaging and MRI, and measure electric field gradients in neuronal activity. In 2022, Shengran Lin delved into the domain of electric-field noise detection on the surface of the diamond, employing shallow NV centers as their investigative tools.[134] Their exploration reveals a fascinating phenomenon: the electric-field noise exhibits a power-law frequency dependence, its behavior intricately intertwined with temperature and the activation energies of two-level systems. As temperature ascends from 295 K to 420 K, the noise experiences a gradual increase, but it diminishes swiftly at even higher temperatures. The origin of this noise lies in ensembles of two-level systems characterized by activation energies spanning from 0.3 eV to 0.6 eV, a behavior elucidated by the Dutta-Horn model. Notably, the study indicates the superiority of solid covering material, specifically PMMA, over its liquid counterpart, glycerol, in mitigating electric-field noise on the diamond surface. This superiority is attributed to the presence of surface-modified photonic coupling noise. Overall, this research enhances our comprehension of the underlying factors contributing to surface 1/f electric-field noise and sheds light on the choice of optimal covering materials for augmenting the coherence time of near-surface NV centers.

In a very recent study in 2023 by Jeffrey Neethi Neethirajan, this paper delves into the formidable challenges confronting near-surface negatively charged NV centers, encompassing issues such as charge-state instabilities, diminished fluorescence, and NV coherence time.[135] These impediments significantly hinder the sensitivity of magnetic imaging, a concern that becomes particularly pronounced at the demanding conditions of 4 K and ultrahigh vacuum (UHV). However, the authors present a promising solution by demonstrating the efficacy of in situ adsorption of $H_2O$ on the diamond surface, which offers a pathway toward the partial restoration of shallow NV sensors. They highlight the pivotal role of controlled surface treatments as an indispensable element in implementing NV-based quantum sensing protocols under cryogenic UHV conditions. This research paves the way for improved performance and applications of NV-based quantum sensors in challenging environments.

Quantum cascade lasers (QCLs) exploit quantum mechanical principles, providing coherent and tuneable mid-infrared and terahertz radiation. QCLs are used in spectroscopy to identify specific molecular vibrations, enabling the detection of trace gases, pollutants, and chemicals in diverse environments. Impactful research in 2020 by Yohei Matsuoka was carried out on QCL.[136] It is a remarkable semiconductor device that emits light through electronic transitions between subbands within a heterostructure conduction band, granting it the unique capability to emit light at multiple photon energies. This paper offers a comprehensive insight into the fundamental physics underpinning QCLs, shedding light on their intricate fabrication process and introducing a novel approach to achieving enhanced power and brightness through broad-area QCLs. The paper also delves into the intriguing physics and optics associated with external cavity-tuneable QCLs, which present the fascinating ability to tune the laser's emission wavelength finely. In contrast to conventional diode lasers, which emit light at a fixed energy level, QCLs exhibit the flexibility to emit light at various photon energies, making them a promising candidate for different applications in photonics and spectroscopy.

Cang Wei, in 2020, researched MRI and introduced an efficient and swift approach to fabricate Gd-doped $ZnCuFeS_3$ QDs adorned with surface ligands such as oleyl amine, oleic acid, and mercaptan.[137] Due to their remarkable features, these QDs hold great promise as contrast agents in nuclear magnetic resonance imaging. Notably, they exhibit an adjustable fluorescence spectrum, allowing for versatile imaging applications while maintaining stable chemical properties. This innovative method opens new possibilities for enhancing the precision and versatility of imaging techniques, making it a valuable contribution to medical diagnostics and research.

In 2022 by Zhenjie Wang (Single Photon Avalanche Diode Circuits For Ultra-Violet Imaging Zhenjie Wang, Ivor Fleck, Bhaskar Choubey), a study was done describes the development of a compact CMOS single-photon avalanche diode (SPAD) pixel tailored for ultra-violet imaging. Achieved through a standard high-voltage CMOS process with a dedicated passivation opening for the diode, the pixel boasts dimensions of 35×40 μm and an 8% fill factor. An active quenching circuit is introduced to optimize avalanche quenching and recharge, significantly enhancing speed and reliability. Additionally, incorporating a 9-bit digital counter within each pixel facilitates precise photo-counting operations. This innovative SPAD pixel holds promising potential for various applications in ultra-violet imaging, particularly in scenarios requiring single-photon detection capabilities.

An impactful research by Tingting He in 2022 developed a remarkable device featuring a substantial InGaAs absorption region complemented by an anti-reflection layer. This combination has yielded an impressive quantum efficiency of 83.2%.[138] The device's single-photon performance was meticulously assessed by implementing a quenching circuit, resulting in a remarkable maximum detection efficiency of 55.4%. Even more impressively, this achievement was accompanied by a dark count rate of just 43.8 kHz and an incredibly low noise equivalent power of $6.96 \times 10^{-17}$ W/Hz$^{1/2}$ at a temperature of 247 K. What sets this device apart from previously reported detectors is its ability to maintain a higher single-photon detection efficiency (SPDE) while operating at a more elevated cooling temperature, underlining its promising potential for various applications in the field.

In 2022 by Liu Chen, Avalanche-photodiode-based near-infrared single-photon detectors, notably the InGaAs/InP variant, have become increasingly popular owing to their exceptional sensitivity, rapid response times, and seamless integration capabilities.[139] This paper delves into the latest advancements and diverse applications of InGaAs/InP photodiodes, showcasing their continuous performance enhancements achieved through structural optimization and external quenching circuits. These detectors boast remarkable attributes, including substantial internal gain, heightened sensitivity, swift response, compact form factors, and ease of integration. Furthermore, the study provides a brief overview of alternative near-infrared single-photon detection technologies founded on novel materials and mechanisms, hinting at the promising potential for further breakthroughs in this dynamic field.

In summary, due to their unique quantum properties, QMs bring exceptional capabilities to sensing and imaging technologies. From fluorescence imaging to quantum-enhanced spectroscopy and quantum sensing, these materials drive advancements in diverse fields by enabling susceptible, precise, and versatile detection techniques.

*4.3. Quantum Communication and Information Processing*

The innovative realm of QMs has captivated researchers and scientists, propelling them into uncharted territories of quantum communication protocols and quantum information processing. By harnessing the distinct properties of these materials, such as superposition and entanglement, we stand on the precipice of a revolution in information technology. With their ability to manipulate and transmit data at a quantum level, these materials promise exponentially faster and more secure communication systems and computation methods. As we delve deeper into the intricate interplay between QMs and the forefront of technology, we unlock the potential to redefine human knowledge's limits and reshape modern information science's landscape. In this section, we will explore the usefulness of QMs in quantum communication protocols and information processing and discuss the applications of QMs in quantum cryptography, quantum key distribution, and quantum gates.

Quantum communication and information processing leverage the principles of quantum mechanics to achieve tasks beyond classical systems' capabilities. There exist two primary Quantum Communication Protocols: Quantum Teleportation enables the transfer of quantum information across distances through entangled states, while Quantum Cryptography, exemplified by protocols like BB84, ensures secure communication via quantum key distribution (QKD). Quantum Information Processing employs quantum gates and algorithms, which are crucial for tasks like factorization using Shor's algorithm and pivotal for cryptography. Quantum gates, essential components of quantum circuits, rely on mechanisms like superconducting qubits, advancing quantum computing's practicality. Quantum technologies enable secure communication, ensuring data encryption through quantum keys, thus thwarting eavesdropping with unhackable encryption. QKD extends secure communication over long distances, with potential implications for Internet security, promising unbreakable encryption. Quantum gates facilitate the construction of quantum circuits, which are crucial for information processing tasks and scalability in quantum computing. Research into new quantum mechanisms is imperative for advancing quantum computing towards practical applications, and for this, advancement in QMs research is essential.

An impressive paper by Giordano Scappucci in 2021 [140] shows the significance of high-purity crystalline solid-state materials in advancing quantum information processing, particularly in qubit technologies reliant on spins and topological states. It highlights the potential and challenges associated with semiconductor heterostructures for spin qubits, with GaAs, Si, and Ge as established platforms demonstrating two-qubit logic. Additionally, the abstract emphasizes the promise of topologically nontrivial materials like topological insulator thin films and materials such as PbSnTe for creating topological qubits. It calls attention to the need for advancements in fabrication and characterization techniques to enable integration with Josephson junctions and the development of high-quality QMs. In conclusion, the paper identifies the most promising avenues for advancing qubit technology through enhancements in these material systems, with a strong emphasis on improving material quality and heterostructures to mitigate decoherence and enhance solid-state qubit performance.

In 2022, a study was carried out by Yoann Pelet, introducing an achievement in quantum communication, a metropolitan quantum network utilizing energy-time entanglement-based QKD.[141] This pioneering system spans 50 kilometers, connecting three nodes within Nice via optical fibers. It achieves a notable raw key rate of 40 kilobits per second per pair of channels and adheres to the ITU 100 GHz standard for telecom-grid compatibility. To bring this innovative network to life, the researchers harnessed a high-quality source of energy-time entangled photon pairs and implemented robust clock synchronization techniques, eliminating the need for dedicated communication channels. The post-treatment software facilitated real-time key generation, culminating in successfully establishing secret keys. This research marks a historic milestone by presenting the world's first fully operational entanglement-based metropolitan quantum network, demonstrating its feasibility and remarkable performance in a real-world setting.

QMs are pivotal in advancing quantum communication protocols and quantum information processing. Their unique quantum properties enable secure communication, unbreakable encryption, and robust quantum computers. As research into QMs continues, we can expect significant advancements in quantum technology, with profound implications for various fields, including cryptography, information processing, and beyond.

*4.4. Energy Conversion and Storage*

QMs have emerged as revolutionary components in energy conversion and storage devices, transcending the limitations of classical materials. Their exceptional properties, often harnessed at the nanoscale, enable leaps in efficiency and performance. In energy conversion, QMs like perovskite solar cells exhibit unprecedented light-harvesting capabilities, efficiently converting sunlight into electricity. Similarly, QDs offer tuneable electronic and optical properties, enhancing the efficiency of LEDs and photodetectors. QMs such as graphene and TIs in energy storage contribute to supercapacitors and batteries with enhanced charge storage and faster energy release. Quantum phenomena like entanglement can be leveraged to develop quantum-enhanced devices for even more efficient energy conversion and storage, paving the way for a sustainable energy future. In this section, we will describe the utilization of QMs in energy conversion and storage devices and discuss advancements in quantum dot solar cells, thermoelectric materials, and quantum-enhanced energy storage.

Due to their unique properties, tin oxide QDs ($SnO_2$ QDs) are gaining attention for energy storage. Tin oxide is an eco-friendly semiconductor with stability, and it is found in solar cells, capacitors, batteries, and gas sensors.[142] $SnO_2$ QDs have advantages like high surface area for better reactions, cost-effectiveness, high capacity, fast charging, stability, and compatibility with other materials. In a study, a hybrid of $SnO_2$ QDs and activated carbon showed a capacitance of 222 $Fg^{-1}$ and high energy/power densities.[143] This composite holds promise for portable electronics and energy storage systems.

Cadmium sulfide (CdS) QDs possess a large surface area and low band gap (2.42 eV), making them vital II–IV semiconductors.[144] Their exceptional electrochemical, optical, and fluorescence properties are ideal for pharmaceutical analysis and element detection. CdS QDs ' effectiveness is influenced by their size and structure, prompting research into proper forms and dimensions. Various methods, like sol-gel, hydrothermal, organic solvent, and microwave processes, produce CdS QDs with different properties. Smaller CdS QDs exhibit improved capacitance and cyclic stability. Recent studies highlight CdS QDs' potential as electrode materials in supercapacitors (SCs) due to their high capacitance, cycling stability, and rapid charge-discharge rates. Organometallic perovskites and CdS QDs were used in layers for thin-film electrochemical capacitors, enhancing cycling properties and energy densities. Challenges include CdS QDs ' low conductivity and toxicity, prompting strategies like carbon-based hybridization and non-toxic alternatives. While CdS QDs hold promise for SC electrodes, further research is needed to address challenges and fully unlock their potential, potentially through innovative synthesis methods, hybridization, and exploring non-toxic alternatives.[145,146]

QDs of $MoS_2$ have caught the attention of researchers due to their unique properties, like light interactions and electrical behaviour. [147] These dots have various potential applications in sensors, energy, and healthcare. They are small and emit controllable light, making them useful for biosensors and imaging.[148] Scientists have found different ways to make $MoS_2$. They used techniques like UV-Vis's absorbance, luminescence, and microscopy to study the dots' size and appearance. Compared to other $MoS_2$ structures, these dots showed better performance in terms of their electrical behavior. They had a more significant potential range and higher capacity. A device using these dots also showed impressive stability and retained its capabilities even after many use cycles.

Recent advancements in 2D layered inorganic materials, like monolayer and few-layer sheets of tungsten disulfide ($WS_2$) and other dichalcogenides, are currently under close investigation.[149] $WS_2$, with its unique trigonal prismatic structure, falls under transition metal dichalcogenides (TMDs) and shares a layered arrangement like graphite.[150] Due to its potential applications in solid lubricants, lithium batteries, bioimaging, hydrogen evolution reaction (HER), and more, researchers are actively exploring the controlled production of innovative and superior $WS_2$ micro-nanomaterials, sparking widespread interest.

The porous graphene sheet combined with attached GQDs forms a GQDs/Gr hetero junction, enhancing ionic diffusion and active sites for ion storage. At the same time, AL contributes to activated carbon with a dense pore structure. The hybrid supercapacitor works up to a 1.4 V voltage (as shown in the charts) and exhibits specific capacitances of 9.8, 8.5, 7.28, 6.45, and 5.85 F cm$^{-3}$. The CV curves and GCD profiles reveal pseudocapacitive charge storage alongside EDLC. S, N-GQDP$_2$ nanocomposite excels with an estimated 645 Fg−1 Csp, outperforming pure PANI, S, N-GQDP$_1$, and S, N-GQDP$_3$ samples at 177, 224, and 134 Fg−1, respectively.[151] Heteroatom-doped S, N-GQDs exhibit enhanced capacitance due to sulfur and nitrogen trapping electrolyte ions. The incorporation of heteroatoms boosts conductivity. Similar enhancements are seen with $CoMoO_4$ nanosheets modified with chitosan. Combining pseudo-capacitor and quasi-zero-dimensional carbon materials enhances energy storage through their high surface area, conductivity, and stability. Further research is necessary for optimization. Solution- and solid-type electrolytes employ GQDs with acid oxygen groups for improved supercapacitor performance.[152–154] Neutralizing acidic GQD groups enhances capacitance efficiency and rate capability. The storage capacity of nanocrystals and SP-derived CDs in an $H_2SO_4$ electrolyte increases with electrolyte amount, yielding 643 $mFcm^{-2}$ due to reduced resistance and better wettability.

Various characterization techniques were employed to study synthesized graphene QDs (GQDs). Electrochemical experiments were used to explore GQDs' supercapacitive properties.[155] An electrode made from GQDs demonstrated a specific capacitance (SC) of 257 $F\ g^{-1}$ at 3 $A\ g^{-1}$ current density, retaining 96% capacitance after 3000 cycles. GQDs were also used as active material in a flexible symmetric supercapacitor (SSC) with promising outcomes even after 1000 cycles. The SSC exhibited a rectangular cyclic voltammogram (CV) shape, indicative of reversible ion adsorption-desorption and a dual-layer capacitive nature. Galvanostatic charge-discharge (GCD) tests displayed symmetrical triangular curves with minimal internal resistance drops, confirming high-rate capabilities. The SC values at varying current densities were calculated, highlighting the devices' resilience to high current densities. Stability tests further validated the long-term performance.[156] Additionally, GQDs were employed as solid-state electrolytes, offering stability and compatibility with various electrode materials. The work extended to transparent and flexible micro-supercapacitors using GQDs, exhibiting remarkable transparency, energy storage, quick relaxation time, cycle retention, and stability.[157,158] The synergy between $MoS_2$ QDs and PANI in a hybrid conductive xerogel enhanced capacitance and stability.[159] The study demonstrated the potential for novel energy storage devices with wide-ranging applications.

Demand for energy storage devices has spurred the need for cost-effective QDs production, focusing on their potential applications in batteries and supercapacitors.[147] These applications heavily rely on the nature of electrode materials, where carbonaceous materials, particularly graphene and its nano-meter-sized derivatives known as GQDs, have exhibited excellent electrochemical properties due to their structure.[160] GQDs offer unique attributes such as electron-hole pairs for conductivity, functionalization potential, and wettability for improved electrochemical properties. QDs also act as templates for distinct structures and can be incorporated through various methods. Recent studies highlight the impact of quantum dot morphology on the electrochemical performance of composite materials.[161] For instance, CdS QDs with rod-like morphologies displayed enhanced charge storage capacity and cycling stability, a crucial consideration in organic polymer composites for supercapacitor applications. Additionally, the synthesis of quantum dot composites as anode materials for energy storage devices has shown promising results, such as $TiO_2$ QDs anchored on graphene for rapid and stable lithium/sodium storage.[162] These innovative advances, along with ongoing research, hold the potential to shape the future of sustainable and renewable energy sources, addressing pressing environmental challenges and transforming the energy landscape.

A very recent and very effective study by Cheng Zeng in 2023 [163] on the potential of GQDs as interfacial engineering materials for perovskite solar cells (PSC) underlining their appeal due to their low toxicity and superior charge mobility when compared to other metallic-based

QDs within semiconductor applications. However, it also raises a crucial concern about the emergence of structural defects attributed to the use of GQDs, leading to the formation of shallow trap states that contribute to decreased PSC performance. To shed light on this issue, the paper employs thermally stimulated current (TSC) and density-voltage (J-V) plots as analytical tools, offering valuable insights into the influence of structural defects and trap states on PSC efficiency. To mitigate these challenges, the paper proposes implementing a carefully controlled fabrication process, thus offering a path forward to harnessing the potential of GQDs while minimizing their associated trade-offs in PSC technology.

Another recent study of 2023 by Dianli Zhou on indium oxide QDs ($In_2O_3$) was successfully incorporated into a $MAPbI_3$ film, effectively addressing the issue of low film quality caused by inherent defects.[164] Introducing $In_2O_3$ QDs led to a remarkable improvement in film quality by suppressing these defects. The resulting $MAPbI_3$:$In_2O_3$ film exhibited optimized time-resolved photoluminescence component ratios, which indicated enhanced carrier dissociation and improved transport efficiency. When this advanced film was employed in the fabrication of solar cells, it yielded significant improvements in performance metrics. Specifically, the solar cells using the $MAPbI_3$:$In_2O_3$ film demonstrated higher fill factor, short circuit current density, and power conversion efficiency than those using the $MAPbI_3$ film alone. These findings pave the way for developing more efficient and reliable solar cell technologies, with the potential for significant advancements in renewable energy generation.

In 2023 by Yichen Wang in the domain of advanced photodiodes and solar cells.[165] This study explores the design and performance of a 2D $ReS_2$/$MoS_2$ semi-vertical heterojunction photodiode with a unilateral depletion region, showcasing remarkable attributes such as self-powered capability, a broad photovoltaic response, high responsivity, power conversion efficiency, and excellent detectivity. Moreover, this photodiode exhibits an impressively rapid response time and a detectable signal cut-off frequency, rendering it highly suitable for demanding photodiode applications. The 2D $ReS_2$/$MoS_2$ semi-vertical heterojunction photodiode's outstanding performance makes it a promising candidate for various high-performance photodiode applications. Furthermore, incorporating photosensitive PbS QDs into the heterojunction further amplifies its sensitivity, resulting in a significant 21% enhancement in device responsivity under 532 nm illumination. This underlines the immense potential of $ReS_2$/$MoS_2$ heterostructures for utilization in advanced photodiodes and solar cells.

Another very recent study in 2023 by Nanxin Fu introduces a novel photodetector device that capitalizes on a hybrid dimensional heterostructure incorporating multiwall carbon nanotubes (MWCNTs) and multi-layered $MoS_2$.[166] This innovative approach enables the photodetector to achieve broadband detection capabilities from visible to near-infrared light. The results are auspicious, with the device demonstrating exceptional responsivity, detectivity, and external quantum efficiency. Specifically, the photodetector exhibited a remarkable responsivity of $3.67 \times 10^3$ A/W at a wavelength of 520 nm and an impressive 718 A/W at 1060 nm. These values were accompanied by detectivity (D*) figures of $1.2 \times 10^{10}$ Jones at 520 nm and $1.5 \times 10^9$ Jones at 1060 nm. Moreover, the external quantum efficiency (EQE) values were strikingly high, measuring approximately $8.77 \times 10^5$ % at 520 nm and $8.41 \times 10^4$ % at 1060 nm. In summary, this research showcases the potential of mixed-dimensional heterostructures in photodetector technology, enabling proficient detection across both visible and infrared spectra. This development paves the way for innovative optoelectronic devices built upon the foundation of low-dimensional materials, marking a notable advancement in the field.

Advancements in energy conversion and storage technologies have greatly influenced the development of next-gen tech. Batteries and supercapacitors (SCs) play a crucial role in cutting-edge electronics. SCs are reliable for real-time applications due to their benefits, such as high power density and long lifespan at lower costs. They are ideal for future energy-focused tech like solar systems, electric vehicles, and wind power. [167] Scientists are working to enhance SCs, and this review highlights QDs and nanocomposites' role in improving SC performance. Effective QD use boosts SC efficiency, and QDs have broad applications in battery materials, medical tech, sensors, and more. The review zeroes in on QD-based electrode materials and flexible devices, showcasing their potential impact.

*4.5. Spintronics applications and topological superconductors*

TIs represent a fascinating material class that has garnered significant attention due to their unique electronic properties. These materials exhibit an intriguing behaviour wherein their bulk is insulating, yet their surface states conduct electricity robustly, making them resistant to scattering and impurities. This topological protection arises from nontrivial band topology and symmetrical constraints. This section will provide an in-depth look at TIs and their potential applications in quantum computing and spintronics. We will also discuss topological materials' role in achieving robust and efficient electronic states.

Topological superconductors possessing surface metal and interior superconducting states feature Majorana fermions and hold promise for faultless quantum computing. Magnets with topological properties exhibit chiral edge states, merging magnetism and topology, potentially revolutionizing low-energy-loss electronic devices. TIs, showcasing unique behaviours at their surface or edge states, permit electron flow peculiarly and are protected from scattering by time-reversal symmetry. This remarkable property arises from the intricate interplay of quantum mechanics and topology, a branch of mathematics dealing with invariant space properties under continuous deformations. Unlike conventional insulators, they allow electron flow at their surfaces while maintaining insulating properties internally, holding immense potential for quantum computing and spintronics. Quantum computing, leveraging quantum mechanics principles, demands robust platforms for quantum bits or qubits, where TIs offer stable qubits that are less prone to environmental decoherence, which is crucial for building scalable quantum computers. In spintronics, focused on utilizing electron spin for information storage and processing, TIs combine electron spin with a motion to create spin currents exploitable for novel device functionalities, offering energy-efficient computing and data storage solutions. Topological materials facilitate robust and efficient electronic states that are immune to external perturbations or localized defects and are critical for reliable and fault-tolerant devices. Various growth methods exist for topological materials.

Angle-resolved photoemission spectroscopy (ARPES) plays a crucial role in uncovering and understanding topological phases by revealing nontrivial electronic structures.[168]

Additionally, ARPES helps measure topological invariants that classify different phases and study the impact of external forces. A notable example is the topological magnet $CoSn_2S_2$[169], which displays fascinating magnetic properties and potential applications in storage. The quantum anomalous Hall effect (QAHE) is significant for low-energy electronics and topological superconducting quantum computing, and efforts to enhance QAHE continue with high-throughput calculations predicting potential materials. These advances open doors for high-temperature QAHE materials and innovative spintronics devices, harnessing properties for diverse applications like information storage and charge transport. Topological states have been studied before, especially in topological superconducting materials.

A physical property measurement system (PPMS) is a helpful tool to identify superconductivity in these materials.[23] This transition is typically seen in temperature-dependent resistance, magnetization, and heat capacity. Heat capacity measurement at varying temperatures is a reliable way to analyze superconductor properties. There are additional methods to characterize topological superconducting materials, but they need to be more detailed here due to space limitations. Taking $PbTaSe_2$ as an example, it has topological surface states.[170] These surface states possess a distinct isolated band with helical spin texture due to their hybridization with bulk states and spin-momentum locking. This gives rise to helical topological superconductivity. The combination of topology and superconductivity results in a new quantum state, a topological superconductor, which is crucial for quantum computing. Overcoming errors caused by decoherence is a significant challenge in quantum computing, and topological quantum computing can solve this.

Majorana zero-energy modes on the surface of topological superconductors offer high fault tolerance for topological quantum computation. Anisotropic Majorana-bound states have been observed, opening avenues for novel heterostructures.[171] Topological superconductivity has been detected in various systems, with nanowires showing promise. Majorana fermions might emerge from coupling superconductors with spin-orbit-coupled semiconductor nanowires.[172] These advancements hold potential for stable qubit modules in quantum computing and necessitate further research and understanding of the physics involved.

In 2023, K. Imamura conducted a study [173]. In this study, the researchers delved into the intriguing question of the origin of the planar thermal Hall effect within the Kitaev magnet α-RuCl3, a material thought to potentially host Majorana fermions and non-Abelian anyons, vital components of Kitaev spin liquids. Previous observations of an unusual planar thermal Hall effect in α-RuCl3 had stirred controversy over whether it stemmed from Majorana fermions or bosonic magnons. The authors conducted meticulous low-temperature measurements of specific heat and thermal Hall conductivity to resolve this ambiguity, employing in-plane field rotations as a discriminating factor. Their findings showcased a closure of the low-energy bulk gap and a sign reversal of the Hall effect along the honeycomb bond direction. They offered conclusive evidence favoring the Majorana-fermion origin of the thermal Hall effect in α-RuCl3. These results bolster the notion that α-RuCl$_3$'s field-induced quantum-disordered state harbors Majorana fermions and non-Abelian anyons, solidifying its status as a promising candidate for Kitaev spin liquids. This research successfully settles the long-standing debate surrounding the thermal Hall effect's origin, definitively supporting the Majorana-fermion hypothesis based on compelling experimental observations.

In 2023, C. W. J. Beenakker [174] underlined the significance of the method of tangent fermions in discretizing the Hamiltonian of TIs and superconductors. This method stands out for its exceptional ability to safeguard the topological protection of massless excitations, whether Dirac or Majorana fermions. Tangent fermions operate within a 2+1-dimensional lattice framework with a tangent dispersion, sidestepping the problematic fermion doubling lattice artifact that could otherwise jeopardize the system's topological integrity. Although the resulting discretized Hamiltonian exhibits nonlocal characteristics, it can be transformed into a localized generalized eigenproblem, ensuring its practicality in both spatial and temporal domains. Moreover, the paper delves into various applications of tangent fermions, encompassing phenomena such as Klein tunneling, resistance to localization by disorder, the anomalous quantum Hall effect, and the intriguing concept of the thermal metal involving Majorana fermions. Altogether, the method of tangent fermions emerges as an auspicious approach for discretizing the Hamiltonian of TIs and superconductors while upholding their topological protection and preserving the fundamental symmetries inherent in the Dirac Hamiltonian.

In 2022, Martin did an impressive review on the interface of organic and inorganic for spin injection using carbon nanotubes and graphene for spintronics applications.[175] Another domain of quantum technology called spintronics adds spin quantum degrees of freedom to CMOS devices. Since the discovery of giant magnetoresistance in 1988 by Fert[176] and Grunberg[177], spintronics has transformed everyday gadgets like hard drives and magnetic memories. This review focused on Carbon-based nanomaterials like nanotubes, graphene, and molecules, which are physically and chemically compatible. Their high electrical mobilities, weak spin dispersion, and hybridization lead to strong spin polarizations that meet the critical requirements of spintronic devices. In this regard, the first technological endeavor to integrate nanomaterials for spintronics shows superior spin diffusion lengths. They focus on material, physical, and chemical compatibility to achieve remarkable performance in the next generation of spintronics devices. This novel approach needs advances in surface chemistry, spin transport physics, and device theory, as detailed in this article. Ongoing research is breaking new ground, with advancements like all-spin-logic circuits and neuromorphic chips nearing fruition. Carbon nanostructures like molecules, graphene, and nanotubes play a vital role. Recent experiments emphasize the significance of calibrated tunnel barriers in graphene-based spin valves, boosting spin signals and diffusion length in non-magnetic materials. These developments highlight spintronics' dynamic nature and potential to revolutionize modern electronics.

Another study in 2021 on spintronics by Jun-Wen Xu introduced a radio-frequency signal that produces spin dynamics in a spintronic resonator, which the spin-diode effect detects.[178] Ferromagnetic metals and spin torque are used in such resonators. QMs can enable new technologies, such as transition metal oxides with phase transitions that give spintronic resonators hysteresis and memory. It highlights the impact of Ni, Permalloy ($Ni_{80}Fe_{20}$), and Pt layers over $V_2O_3$; the first-order phase transition causes systematic resonance response alterations and hysteretic current control of the ferromagnetic resonance frequency. A DC current can locally change the state of the output signal to vary systematically. These findings show neuromorphic computing-relevant spintronic resonator functions. The study reveals systematic alterations in resonance response and hysteretic current control of ferromagnetic resonance frequency. Moreover, it demonstrates the manipulation of output signal via state changes in $V_2O_3$ induced by a DC current. Temperature significantly influences resonance amplitude, while structural phase transitions in $V_2O_3$ reshape ferromagnetic resonance conditions. Emphasizing the quantum material spintronic resonator's memory retention, this research underscores its remarkable and distinctive attributes.

A study by Swapnali Makdey in 2020 shows a novel and efficient approach for designing magnetic tunnel junctions (MTJs) structures utilizing molybdenum disulfide $(MoS_2)$/GQDs/$MoS_2$.[179] They create a device that exhibits superior efficiency coupled with low power consumption. With 2D $MoS_2$ as ferromagnetic electrodes and GQDs acting as barriers, the MTJ tunnel magnetoresistance (TMR) reached 1450% at zero bias voltage. Investigating spin relaxation and magnetization relaxation further extended the spin's lifetime. The proposed MTJ structure for spintronics demonstrated notable effectiveness, underscored by comparative analysis with existing MTJ designs.

A recent study on MTJs in 2022 by Zhan was carried out to explore the potential of 2D MnBi2Te4 by investigating their spin-dependent electronic and transport properties.[180] In this study, they used the first principles quantum transport simulations. They designed the MTJs based on 2D van der Walls layered $MnBi_2Te_4$ and observed that by increasing the thickness of the $MnBi_2Te_4$ layers within the MTJs, the research found significant enhancements in spin polarization and TMR ratios at the Fermi level. TMR ratios of up to 100% and 500% were observed for varying thicknesses of $MnBi_2Te_4$ layers, with a remarkable increase to 500% and 4000% when accounting for the spin-orbit coupling (SOC) effect. These findings represent the potential of $MnBi_2Te_4$ -based MTJs for advancing spintronic device performance, suggesting promising avenues for further investigation and development in this field.

So far, we summarized TIs and spintronics applications, which stand at the intersection of quantum mechanics, topology, and condensed matter physics, offering the potential to revolutionize quantum computing and spintronics applications. Their ability to host protected electronic states holds promise for revolutionizing the landscape of computing and information processing. As researchers delve deeper into the intricate properties of these materials, the future of technology may be shaped by their remarkable properties and capabilities.

*4.6. QMs for Quantum Computing*

The exploration of QMs has emerged as a pivotal endeavor in advancing quantum computing platforms. QMs play a crucial role in the development and advancement of quantum computing. This paradigm-shifting technology leverages the principles of quantum mechanics to perform complex calculations and solve problems currently intractable for classical computers. QMs provide the physical platforms upon which quantum bits (qubits), the fundamental units of quantum information, are realized and manipulated. These specialized materials possess unique and often exotic properties that can be harnessed to manipulate and control quantum states, enabling the development of more efficient and powerful quantum bits or qubits—the fundamental building blocks of quantum computers. Researchers are delving into the intricate behavior of QMs, leveraging their quantum coherence, superconductivity, and topological properties to create stable and reliable qubits. By focusing on the synthesis, characterization, and utilization of these materials, scientists aim to unlock the full potential of quantum computing, revolutionizing computation and opening doors to unprecedented computational capabilities with profound implications across diverse fields of science and technology. This section focuses on QM's role in developing quantum computing platforms and discusses qubit implementations using superconducting circuits, trapped ions, and topological qubits.

QMs are fundamental for constructing qubits, the core components of quantum computers, which can exhibit quantum properties crucial for their operation. QMs allow for scalable integration of qubits into larger systems, enabling the creation of multi-qubit systems interconnected for complex calculations. For example, superconducting materials facilitate the creation of superconducting qubits, harnessing quantum behaviors like superposition and entanglement. Superposition enables quantum computers to process vast amounts of information simultaneously, while entanglement is vital for operations surpassing classical computing capabilities. QMs also allow the implementation of quantum gates, which are fundamental for manipulating qubits and executing computations in parallel. Quantum error correction, essential for reliable quantum computing, utilizes QMs to create fault-tolerant qubits that are less sensitive to environmental disturbances. Specific QMs, such as superconducting qubits, play critical roles in specialized quantum computing approaches like quantum annealing, which exploits quantum tunneling effects for optimization problems. Researchers explore how quantum properties of materials can inspire novel algorithms for more efficient problem-solving leveraging quantum parallelism and interference.

An awe-inspiring study In 2021 by Oreg Yuval [181] explored a quantum computing device comprising a carbon nanotube, superconducting substrate, and magnet, with a particular emphasis on the spin-triplet aspect and the longitudinal magnetic field. Furthermore, the study delves into an alternative quantum computing setup involving multiple superconducting substrates and a non-superconducting structure exhibiting robust spin-orbit coupling interactions, highlighting the significance of phase variations among substrate order parameters. The paper's conclusion affirms the feasibility of constructing a quantum computing device using a carbon nanotube, superconducting substrate, and longitudinal magnetic field, with a keen focus on the spin-triplet element. Additionally, it posits that an alternative quantum computing apparatus can be realized by utilizing multiple superconducting substrates and a non-superconducting structure with potent spin-orbit coupling interactions, stressing the importance of phase distinctions within substrate order parameters.

A review by Vincenzo Lordi in 2021 [182] discussed the immense potential of quantum computers over classical counterparts; the presence of noise and material imperfections has emerged as a formidable obstacle. The ongoing progress in materials synthesis, characterization, and modeling is a pivotal cornerstone in surmounting these challenges, thereby unlocking the full capabilities of quantum computing. In recent years, these advancements have catalyzed exciting breakthroughs in the noisy intermediate-scale quantum (NISQ) realm, where the delicate balance between preserving the excellence of single qubits and facilitating high-fidelity qubit interactions is paramount. As we look ahead, it becomes evident that the continued advancement in the synthesis, characterization, and modeling of materials for quantum computing will remain integral to shaping the future landscape of quantum technology.

An impactful review was carried out in 2022 by Axel Hoffmann [183]. In this review, he focused on Neuromorphic computing, which relies on QMs to efficiently process extensive data volumes, as these materials possess distinctive properties that facilitate energy-efficient hardware implementations of neuromorphic concepts. These materials display robust correlations, leading to profoundly non-linear responses that can be effectively utilized for both short- and long-term plasticity and data classification. In conclusion, the utilization of QMs holds significant promise for enabling energy-efficient neuromorphic computing at the hardware level, thanks to their inherent characteristics, which empower them to offer valuable solutions in data processing and artificial intelligence.

In a most recent study in 2023 by Benjamin A. Jackson, he introduced diamond-like structures where Li+ ions and diamines replace carbon atoms and C-C bonds, respectively, creating novel materials capable of hosting diffuse electrons around each lithium tetra-amine center.[184] These materials display either metallic or semiconductor properties, contingent upon the length of the diamine chain. Our gas-phase calculations accurately predict the properties of proposed crystalline Li-diamine materials, offering the potential for further development and insights. Through spin-polarized and unpolarized calculations across various hydrocarbon sizes, we reveal valuable information about their geometrical and electronic band structures, spin density contours, and density of states. These materials hold promise for applications such as redox reactions and quantum computing, where diffuse electrons can be harnessed as qubits. Our future endeavors will tailor the hydrocarbon backbone to control electron association for precise quantum computing and propose materials suitable for selective redox catalysis.

We summarized that QMs provide the physical basis for creating, manipulating, and controlling qubits in quantum computing systems. Their unique quantum properties, such as superposition and entanglement, are essential for performing quantum operations that enable quantum computers to tackle complex problems in areas such as cryptography, optimization, material science, and artificial intelligence. The design, engineering, and study of QMs are thus at the forefront of advancing quantum computing technology.

Table 3: A compilation of review papers highlighting recent advancements in QMs applications.

| S.No. | Title of Paper | Ref. |
|---|---|---|
| 1. | Gate-Controlled Quantum Dots Based on 2D Materials | [185] |
| 2. | The Promise of Soft-Matter-Enabled Quantum Materials | [186] |
| 3. | Organic quantum materials: A review | [187] |
| 4. | Quantum Spin Liquids from a Materials Perspective | [188] |

| 5. | Benchmarking Noise and Dephasing in Emerging Electrical Materials for Quantum Technologies | [189] |
|---|---|---|
| 6. | Quantum-Engineered Devices Based on 2D Materials for Next-Generation Information Processing and Storage | [190] |
| 7. | Graphene/Quantum Dot Heterostructure Photodetectors: From Material to Performance | [191] |
| 8. | Recent advances in topological quantum anode materials for metal-ion batteries | [192] |
| 9. | Materials for Silicon Quantum Dots and their Impact on Electron Spin Qubits | [193] |
| 10. | Quantum Sensing for Energy Applications: Review and Perspective | [194] |
| 11. | Layered materials as a platform for quantum technologies | [195] |
| 12. | Heterostructures of 2D materials-quantum dots (QDs) for optoelectronic devices: challenges and opportunities | [196] |
| 13. | Hexagonal Perovskites as Quantum Materials | [196] |
| 14. | Micro-Light-Emitting Diodes Based on InGaN Materials with Quantum Dots | [197] |
| 15. | Topological quantum materials for energy conversion and storage | [198] |
| 16. | A Review on Quantum Dot Light-Emitting Diodes: From Materials to Applications | [199] |

Summarizing all applications of QMs in a review paper is a challenging task. To aid in this understanding, we have included a helpful reference in Table 3, where we cite recent review papers that delve into the diverse applications of QMs. This table is a valuable resource for readers seeking a more comprehensive insight into the subject. This section explores numerous applications and critical research prospects related to QMs.

## 5. Current challenges in QMs

QMs, with their unique electronic and optical properties arising from quantum mechanical effects, hold great promise for many applications, including quantum computing, electronics, energy storage, etc, which is discussed in detail in section 3. However, After reviewing the literature, they also come with several challenges that researchers are actively addressing. Here are some of the current challenges in the field of QMs:

*5.1. Synthesis and Scalability*: Producing high-quality QMs consistently and at scale can be challenging. Many QMs are delicate and sensitive to environmental conditions during synthesis. Researchers need to develop reliable methods for large-scale production to make these materials practical for real-world applications.

*5.2. Characterization and Imaging techniques*: QMs often exhibit complex behaviors that are not fully understood. Comprehensive characterization, imaging techniques, and theoretical models are needed to unravel the intricate quantum phenomena. This includes studying the emergence of novel electronic phases and understanding the role of defects in these materials.

*5.3. Stability and Environmental Sensitivity*: QMs can be highly sensitive to external factors such as temperature, pressure, and humidity. Ensuring their stability and performance under various conditions is critical for practical applications. Researchers are working on encapsulation and protection strategies to mitigate these effects.

*5.4. Integration into Devices*: Integrating QMs into functional devices, such as transistors or sensors, presents significant challenges. Interfaces and compatibility with existing technologies must be addressed to harness the unique properties of QMs effectively.

*5.5. Quantum Control and Manipulation*: QMs are often used in quantum information processing and quantum computing. Achieving precise control and manipulation of quantum states within these materials is a substantial challenge, as it requires overcoming issues related to coherence times, noise, and error correction.

*5.6. Scalability in Quantum Computing*: While QMs show promise for quantum computing, scaling up quantum systems to a practical level with many qubits remains a formidable challenge. Overcoming issues related to quantum decoherence and error correction is crucial for realizing the potential of quantum computers.

*5.7. Safety and Ethical Concerns*: As QMs and quantum technologies advance, ethical and safety concerns must be addressed. This includes encryption protocols vulnerable to quantum attacks, potential societal impacts, and regulatory considerations.

*5.8. Commercialization and Industry Collaboration*: Bridging the gap between fundamental research in QMs and their commercialization is essential. Collaborations between academia and industry are needed to translate research findings into practical applications effectively.

*5.9. Standardization*: Establishing standards for synthesizing, characterizing, and evaluating QMs is crucial for consistency and reproducibility in research and development.

*5.10. The challenge is Quantum Coherency and Decoherency*: QM devices rely on maintaining the delicate quantum states that imbue them with their extraordinary properties. However, these states are vulnerable to external influences that can disrupt quantum coherence through a process known as decoherence. Shielding devices from environmental noise, temperature fluctuations, and electromagnetic interference remains a formidable challenge.

Addressing the multifaceted challenges QMs poses demands a concerted effort from various disciplines, including physicists, chemists, materials scientists, and engineers. These materials often exhibit intricate quantum phenomena and unique properties, necessitating a deep understanding of their underlying physics, precise chemical synthesis techniques, comprehensive characterization methodologies, and the engineering skills to integrate them into practical devices. QMs hold immense promise in revolutionizing electronics, energy storage, and quantum computing. Thus, overcoming the obstacles associated with their synthesis, manipulation, and integration into devices is not merely a scientific endeavor; it is essential for harnessing their full potential and realizing their transformative impact on various technological domains. Successful interdisciplinary collaboration is the key to surmounting these challenges and ushering in a new era of innovation and discovery.

**6. Discussions and Future Prospects**

QMs have emerged as a fascinating and up-and-coming field of research with profound implications for various scientific disciplines and technological applications. This discussion will delve into the current state of QMs research, as well as its significance and prospects. QMs Research is one of the most demanding fields in the world. By Discovery and Exploration of QMs, researchers have made remarkable strides in discovering and characterizing various QMs over the past few decades. These materials exhibit unique properties, often resulting from their nanoscale or 2D nature, as well as the presence of strong quantum effects. QMs have diverse Classes, which is discussed in the paper. QMs encompass many classes, including TIs, high-temperature superconductors, 2D materials like graphene, and many more. Each class offers its own set of fascinating phenomena and potential applications. Advances in experimental techniques, such as high-resolution electron microscopy and spectroscopy, have enabled researchers to study QMs at the atomic and electronic levels, providing deeper insights into their behavior. Theoretical models and computational simulations have played a crucial role in understanding and predicting the behaviour of QMs. Quantum field theories and density functional theories have been instrumental in advancing our knowledge in this field.

The significant prospects of QMs in Various Applications can be explained very briefly. QMs hold the potential to serve as qubits in quantum computers, enabling exponentially faster calculations for problems that are currently intractable for classical computers. Their inherent quantum properties make them promising candidates for quantum information processing. QMs can be used to create ultra-sensitive sensors and imaging devices. These technologies have medical diagnostics, mineral exploration, and national security applications. QMs' unique electronic and thermal properties can contribute to developing more efficient energy generation and storage systems, including advanced photovoltaics, thermoelectric generators, and supercapacitors. QMs could play a crucial role in developing secure quantum communication systems. Quantum cryptography and quantum key distribution protocols could be realized using these materials. QMs insights can revolutionize materials science by offering new avenues for designing novel materials with tailored electronic, magnetic, and optical properties, enabling innovations in electronics to photonics. QMs have the potential to revolutionize electronics. They can allow the development of ultra-fast, low-energy-consumption electronic devices, paving the way for the next generation of computing and data processing. Some QMs exhibit superconductivity at relatively high temperatures, holding the key to efficient energy transmission and storage. They have the potential to transform power grids and transportation systems. QMs are crucial for quantum computing and quantum information processing. They can provide the qubits needed for quantum computers, which could solve complex problems exponentially faster than classical computers. Topological QMs possess unique topological properties that make them resilient against defects and disorder. These materials could find applications in robust electronic and photonic devices.

Future research will likely focus on designing and engineering QMs with specific properties for targeted applications. This may involve the creation of customized 2D materials or synthesizing exotic materials with desired quantum characteristics. Exploring advanced fabrication methods, such as molecular beam epitaxy and atomic layer deposition, can enable precise control over material structure and composition, enhancing reproducibility and scalability. QMs may play a pivotal role in solving energy challenges. Developing superconductors with higher transition temperatures and improved energy storage materials could transform the energy landscape.

Advancements in QMs will continue to drive progress in quantum computing. Researchers will work on developing more stable and scalable qubits, bringing quantum computers closer to practical use. To counter the effects of decoherence, integrating quantum error correction protocols into device designs can enhance the robustness of quantum states. Quantum error correction codes and fault-tolerant architectures can extend the coherence times of quantum systems.

The prospects for developing machine learning algorithms to tackle open problems in QMs are undeniably promising. As the field of QMs continues to expand, the complexity of materials and their properties has grown significantly, posing formidable challenges to conventional computational methods. Machine learning, with its capacity to analyze vast datasets and identify hidden patterns, offers a transformative approach to predicting and understanding the behavior of QMs. By harnessing the power of artificial intelligence, researchers can expedite the discovery of novel materials with exceptional electronic, magnetic, or optical properties, paving the way for wondering applications in electronics, energy storage, and quantum computing. The ongoing discussions and collaborations within the scientific community regarding the development and application of machine learning in this field are crucial for advancing our knowledge of QMs and unlocking their full potential.

The prospects in designing and engineering exotic phenomena in van der Waals QMs hold tremendous promise for fundamental science and technological advancements. We can anticipate impactful discoveries and innovative applications as we delve deeper into understanding the intricate interplay of electronic, optical, and mechanical properties in these materials. Van der Waals materials offer a versatile platform for tailoring quantum states and exotic behaviour, enabling the development of novel electronic devices, quantum computing technologies, and energy-efficient optoelectronics. This evolving research frontier is poised to unveil many unprecedented phenomena, paving the way for transformative advancements in material science, quantum technologies, and beyond. The ongoing discussions and collaborations in this field will likely stimulate breakthroughs and redefine the boundaries of what is possible in condensed matter physics and engineering.

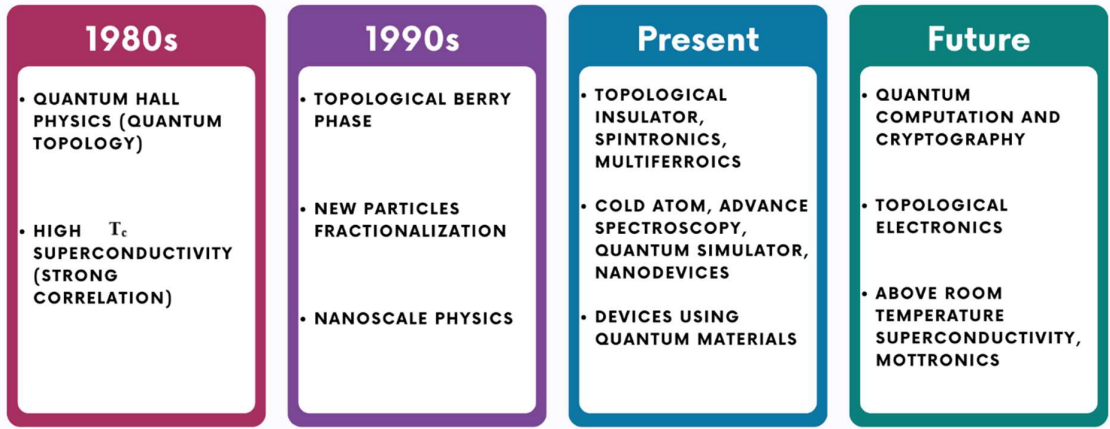

**Figure 9:** Graphic representation summarizes the key concepts, historical developments, and future scientific consequences associated with different QMs.

Exploring new physics in interacting quantum many-body systems holds immense promise for the future of fundamental science and practical applications. As we delve deeper into the complex and intricate world of quantum mechanics, researchers are poised to unlock revolutionary insights into the behavior of matter at the most minor scales. These new frontiers in theoretical physics may pave the way for emerging technologies such as quantum computing and quantum simulations, revolutionizing our ability to solve previously insurmountable problems. Moreover, the study of emergent phenomena in these systems has the potential to unravel mysteries surrounding exotic states of matter, such as TIs and quantum Hall systems, with far-reaching implications for our understanding of the universe. By pushing the boundaries of our knowledge, we can anticipate a future where quantum many-body systems not only provide deeper insights into the fundamental laws of nature but also empower us to harness their capabilities for transformative technological advancements.

Creating hybrid systems that combine the strengths of QMs devices with established technologies can facilitate compatibility. This involves designing interfaces that seamlessly transfer information between quantum and classical components. QMs will continue to be a rich source of fundamental scientific discoveries. Researchers will explore new quantum phases, uncover novel phenomena, and deepen our understanding of quantum physics.

Developing specialized measurement techniques, such as quantum non-demolition measurements and quantum sensing, can enable precise characterization of quantum states without compromising their integrity. QMs can enable ultra-sensitive sensors for various applications, including medical diagnostics, environmental monitoring, and security.

Figure 9 demonstrates that topological currents in QMs promise the future of nonreciprocal electronics and robust quantum computing. They open the door to dissipation-less electronics, magnetoelectronic integrated circuits, high-density storage, superconductors, and quantum computing for energy and information technology infrastructure. These challenges offer revolutionary impacts, though they require further development. Current manifestations, like Mott transitions, high-temperature superconductivity, and colossal magnetoresistance, pave the way for next-generation quantum technologies. We are progressing, but there is more to explore before fully realizing these transformative potentials.

In summary, QMs research represents an exciting frontier with vast potential. It is poised to drive innovation across multiple domains, from electronics to energy to fundamental physics. As our understanding of these materials deepens and technology advances, we can expect to see increasingly transformative applications that will shape the future of science and technology. Significant challenges accompany the integration and commercialisation of QMs devices, but the potential benefits across diverse applications make this pursuit highly promising. By addressing current limitations through innovative approaches and capitalizing on the unique quantum properties of these materials, a future enriched with transformative technologies seems achievable.

## 7. Conclusions

In conclusion, this comprehensive review paper has shed light on the intricate world of QMs, unveiling their pivotal role in reshaping the technological landscape. Throughout this journey, we have delved deep into these materials' unique properties (quantum confinement, strong electronic correlations, and topology and symmetry) and explored their wide-ranging applications across diverse fields. This review highlights the profound and transformative influence of QMs in the ongoing revolution of technology. As we stand on the cusp of unprecedented advancements, it is increasingly clear that the potential of these materials is nothing short of extraordinary. QMs promise ground-breaking innovations in electronics, energy generation, and numerous other domains. We must nurture a persistent spirit of exploration and inquiry to harness their full potential.

The future of QMs research is paramount, as it is critical to unlocking new horizons of innovation and discovery. These materials, with their profound implications and limitless possibilities, are a testament to the remarkable journey that science and technology are poised to embark upon. Our collective responsibility is to continue pushing the boundaries of knowledge and technology, guided by the wondrous possibilities that QMs offer. Through collaboration, dedication, and curiosity, we can harness the full transformative power of these materials and usher in a new era of scientific and technological achievement.

**Statements & Declarations**

**Acknowledgments:** Authors acknowledge Dr. V. Narayanan and Dr. Santosh Mogurampelly, Department of Physics, IIT Jodhpur, India, for providing fruitful comments and suggestions while preparing the manuscript.

**Funding:** Rajat Kumar Goyal received financial support during MTech. from the Ministry of Education, India, during the preparation of this manuscript.

**Competing Interests:** The author declare that they have no conflict of interest.

**Author Contributions:** Rajat Kumar Goyal designed it and wrote the first and final draft of the manuscript.